\documentclass[aps,prd,10pt,superscriptaddress,twocolumn,nofootinbib,dvipsnames,longbibliography]{revtex4-1}

\usepackage[pdftex]{color}
\usepackage[sort&compress]{natbib}
\usepackage[colorlinks=true,linkcolor=purple,filecolor=orange,urlcolor=blue,citecolor=purple,pdftex,plainpages=false]{hyperref}
\usepackage{url}
\pdfoutput=1
\usepackage{ifpdf}
\usepackage[utf8]{inputenc}

\usepackage{color,hhline,psfrag,rotating}
\usepackage{dcolumn}
\usepackage{verbatim}
\usepackage{bm}
\usepackage{slashed}
\usepackage{rotating}
\usepackage{graphicx}
\usepackage[dvipsnames]{xcolor}
\usepackage{amsfonts,amssymb,amsmath}
\usepackage{booktabs}  

\newcommand{\sea}{\text{sea}}
\mathchardef\mhyphen="2D
\DeclareMathOperator{\Tr}{Tr} 

\begin{document}

\global\let\newpage\relax

\title{Renormalising vector currents in lattice QCD 
using momentum-subtraction schemes}

\author{D.~Hatton}
\email[]{d.hatton.1@research.gla.ac.uk}
\affiliation{SUPA, School of Physics and Astronomy, University of Glasgow, Glasgow, G12 8QQ, UK}
\author{C.~T.~H.~Davies}
\email[]{christine.davies@glasgow.ac.uk}
\affiliation{SUPA, School of Physics and Astronomy, University of Glasgow, Glasgow, G12 8QQ, UK}
\author{G.~P.~Lepage}
\affiliation{Laboratory for Elementary-Particle Physics, Cornell University, Ithaca, New York 14853, USA}
\author{A.~T.~Lytle}
\affiliation{INFN, Sezione di Roma Tor Vergata, Via della Ricerca Scientifica 1, 00133 Roma RM, Italy}
\collaboration{HPQCD collaboration}
\homepage{http://www.physics.gla.ac.uk/HPQCD}
\noaffiliation

\date{\today}

\begin{abstract}

We examine the renormalisation of 
flavour-diagonal vector currents in lattice QCD with the aim of 
understanding and quantifying the systematic errors 
from nonperturbative artefacts associated 
with the use of intermediate momentum-subtraction schemes. 
Our study uses the Highly Improved Staggered Quark (HISQ) action 
on gluon field configurations that include $n_f=2+1+1$ flavours 
of sea quarks, but our results have applicability to other quark 
actions.  
Renormalisation schemes that make use of the exact lattice vector 
Ward-Takahashi identity for the conserved current 
also have 
renormalisation factors, $Z_V$, for nonconserved 
vector currents that are free of contamination by nonperturbative 
condensates.
We show this by explicit comparison of two such schemes: that of 
the vector form factor at zero momentum transfer and the RI-SMOM 
momentum-subtraction scheme. The two determinations of $Z_V$
differ only by discretisation effects (for any value of momentum-transfer 
in the RI-SMOM case). 
The RI$^{\prime}$-MOM scheme, although widely used, does not share this 
property. We show that $Z_V$ determined in the standard way in this scheme
has $\mathcal{O}(1\%)$ nonperturbative contamination 
that limits its accuracy. Instead we define an RI$^{\prime}$-MOM
$Z_V$ from a ratio of local to conserved vector current vertex functions 
and show that this $Z_V$ is a safe one to use in lattice QCD calculations.  
We also perform a first study of vector current renormalisation with 
the inclusion of quenched QED effects on the lattice using the RI-SMOM scheme.

\end{abstract}

\maketitle

\section{Introduction} \label{sec:intro}

Lattice QCD is the method of choice for the accurate calculation of hadronic 
matrix elements needed for a huge range of precision particle physics 
phenomenology aimed at uncovering new physics. 
Compelling evidence of new physics in the comparison of experiment 
to the Standard Model (SM) has so far proved elusive, however, and 
this is driving the need for smaller and smaller uncertainties 
on both sides. 
This means that the error bars 
from lattice QCD calculations must be reduced to sub-1\% levels. 
Here we address uncertainties coming from the renormalisation of lattice 
QCD operators to match their continuum QCD counterparts. This 
renormalisation is needed so that the hadronic matrix 
elements of the operators calculated in lattice QCD can be used in continuum
phenomenology. Ideally the uncertainty from the renormalisation factors, $Z$, should 
be much less than other lattice QCD uncertainties (such as statistical errors) 
in the hadronic matrix element calculation.   
 
Defining QCD on a space-time lattice provides an ultraviolet cutoff on the 
theory of $\pi/a$ where $a$ is the lattice spacing. This is a different regularisation 
than that used in continuum formulations of QCD and hence we expect a finite 
renormalisation to be required to match lattice QCD and continuum operators. 
This renormalisation takes account of the differing ultraviolet behaviour in 
the two cases and hence can be calculated as a perturbative 
series in the strong coupling constant, $\alpha_s$, at a scale related to the 
ultraviolet cutoff. 
Lattice QCD perturbation theory is 
notoriously difficult, however, and very few renormalisation constants have been 
calculated beyond $\mathcal{O}(\alpha_s)$ (for an example of a two-loop renormalisation
in lattice QCD perturbation theory see~\cite{Mason:2005bj}). 
It therefore seems clear that this route will not give accurate enough results 
for the future. 

Instead we concentrate here on other approaches that can be 
implemented using results from within 
the nonperturbative lattice QCD calculation. These approaches will typically 
still need to make use of perturbation theory to provide a full matching 
to a preferred continuum scheme such as $\overline{\text{MS}}$, but if this
perturbation theory can be done in the continuum to high order then much improved 
accuracy should be possible.  

At the heart of these nonperturbative-on-the-lattice approaches 
is always the idea that we can  
construct a short-distance operator on the lattice whose leading term 
in an operator product expansion is the operator that we wish to study. 
The matrix elements that 
we calculate on the lattice, and use to determine $Z$, will be dominated 
by those from the leading operator. There will inevitably be contamination, however, from 
subleading terms in the expansion, i.e. higher-dimension operators
multiplied by inverse powers of some scale. 
This means then 
that nonperturbative artefacts 
can enter the determination of $Z$ and these must be understood and controlled 
in order to make use of the $Z$ obtained~\cite{Lytle:2018evc}. 

Here we will study the renormalisation 
factor $Z_V$ associated with the flavour-diagonal vector current that couples to 
the photon. This current is conserved in continuum QCD and has no anomalous dimension. 
Hence we can study 
the lattice QCD determination of $Z_V$ directly, and its dependence on the lattice 
spacing, without having to combine it with a matrix element for the vector current 
determined in lattice QCD. $Z_V$ is a special case of a renormalisation constant 
that can be calculated exactly in lattice QCD, i.e. without the need for 
any continuum perturbation theory and without nonperturbative 
artefact contamination. 
It is important to use a method that allows for such a calculation if we want 
an accurate normalisation.  

It is possible to write down conserved vector currents in lattice QCD and use 
these, knowing that they do not require renormalisation because 
there is an exact vector Ward-Takashashi identity. 
Conserved vector currents are not generally used, 
however, because they are 
complicated objects, especially for discretisations of QCD that are highly improved. 
The removal of tree-level discretisation errors at $\mathcal{O}(a^2)$ from the 
covariant derivative in the Dirac equation requires the addition of operators 
that extend over three links~\cite{Naik:1986bn}. The conserved current then
contains both one-link and three-link terms and this is the case for the Highly Improved 
Staggered Quark (HISQ) action that we will use here 
(see Appendix~\ref{app:cons-curr}). We demonstrate explicitly how the vector 
Ward-Takahashi identity works in this case. 

The HISQ action was
designed~\cite{Follana:2006rc} to have very small discretisation effects 
and this allows its use to study both light and heavy quark 
phenomenology~\cite{Follana:2007uv}. 
Whenever a vector current is needed for phenomenology, however, it is much 
easier to use a nonconserved local (or simple one-link point-split) 
vector current than the conserved one~\cite{Donald:2012ga, Donald:2013pea, Donald:2013sra}. 
This must then be renormalised. 
 
Renormalisation schemes for nonconserved currents 
that make use (not necessarily explicitly) of ratios of matrix 
elements for conserved and nonconserved vector currents 
have a special status because nonperturbative 
contributions from higher dimension operators are suppressed 
by powers of $a^2$. They give renormalisation constants, $Z_V$, 
for nonconserved lattice vector currents that are exact in 
the $a \rightarrow 0$ limit. 
Such a $Z_V$ can then be combined with 
a matrix element of that nonconserved current in 
the lattice QCD calculation and 
the result extrapolated to zero lattice spacing. 
The same answer will be obtained 
in that limit with any such $Z_V$. 

Following the discussion of perturbative matching 
earlier we can think of 
an exact $Z_V$ as consisting of a perturbative series in $\alpha_s$ 
that depends 
on the form of the vector current 
(and also in principle terms arising from small instantons~\cite{Novikov:1984rf} 
or other nonperturbative effects of this kind) 
plus discretisation effects that depend on 
the scheme and vanish as $a \rightarrow 0$~\cite{Vladikas:2011bp}. 
Note that we do not need to know what the perturbative series is; 
the method is completely nonperturbative. 
Which exact $Z_V$ to use is then simply an issue of numerical cost to achieve a 
given uncertainty and/or convenience.  

One standard exact method for renormalising 
nonconserved vector currents in lattice 
QCD is to require (electric) charge conservation i.e. that the vector form 
factor between identical hadrons at zero momentum transfer should 
have value 1. 
Since this result would be obtained for the conserved current, $Z_V$
is implicitly a ratio of nonconserved to conserved current matrix elements 
between the two hadrons. 
This method is numerically fairly costly because it requires the calculation of 
two-point and three-point correlation functions. It can give numerically accurate 
results ($\mathcal{O}(0.1\%)$ uncertainties) 
when averaged over a sufficiently large sample 
(hundreds) of gluon field configurations. 
As above, we expect the $Z_V$ determined from this method (which we will denote 
$Z_V(\mathrm{F(0)})$) to be equal to a perturbative 
matching factor up to discretisation effects. 
This was tested by HPQCD in Appendix 
B of~\cite{Chakraborty:2017hry} for the local vector current made of HISQ quarks. 
Values for $Z_V^{\mathrm{loc}}(\mathrm{F(0)})$ were calculated at multiple values of the lattice spacing 
and gave a good fit to a perturbative expansion in $\alpha_s$ plus discretisation effects, 
constraining the $\mathcal{O}(\alpha_s)$ coefficient to have the known value determined 
in lattice QCD perturbation theory. 

Alternative methods of determining renormalisation factors by defining a variety of  
momentum-subtraction schemes on the 
lattice~\cite{Martinelli:1994ty,Chetyrkin:1999pq,Aoki:2007xm,Sturm:2009kb} 
can produce precise results for $Z$ factors
at lower computational cost. 
However, only some of these schemes are exact for $Z_V$ in the sense defined 
above. 

The momentum-subtraction schemes define $Z_V$ from the ratio of two 
matrix elements calculated between external quark states of large virtuality, $\mu^2$, 
in a fixed gauge. 
Working at large $\mu^2$ is part of the definition of these 
schemes because nonperturbative contributions from 
higher-dimension operators will in general be suppressed by powers 
of $\mu^2$ and not $a^2$ as above.  
A wavefunction renormalisation factor is 
determined from the quark propagator. A vertex renormalisation factor 
comes from an amputated vertex function 
for the vector 
current, on which momentum-subtraction renormalisation conditions have been imposed. 
$Z_V$ is then obtained as the ratio of these two factors, 
with tiny statistical errors 
from a handful of gluon field configurations 
if `momentum sources' are used~\cite{Gockeler:1998ye}. 

The momentum-subtraction scheme known as 
RI-SMOM~\cite{Sturm:2009kb} is constructed around the 
Ward-Takahashi identity and so designed to give $Z_V=1$ for 
the lattice conserved current.  We show explicitly that this is 
true for the HISQ action. 
This means that implementing the RI-SMOM scheme for 
nonconserved currents is equivalent to taking a 
ratio of vector vertex functions for conserved and nonconserved currents. 
We compare the $Z_V$ values obtained in the 
RI-SMOM scheme, $Z_V({\text{SMOM}})$, to 
those from the form factor method for the local vector HISQ current. 
We are able to show that, as expected, $Z_V^{\text{loc}}({\text{SMOM}})$ 
differs from $Z_V^{\text{loc}}({F(0)})$ only by discretisation 
effects so that the two methods 
will give the same answer for physical matrix elements in the continuum limit. 

A popular momentum-subtraction scheme that does {\it not} make use of the 
vector Ward-Takahashi identity is the RI$^{\prime}$-MOM 
scheme~\cite{Martinelli:1994ty,Chetyrkin:1999pq}. 
We show that in this scheme the $Z_V$ values for both the 
conserved and local vector currents are
not exact but have contamination from nonperturbative (condensate) 
artefacts that 
survive the continuum limit.  
To make use of this scheme $Z_V$ must be redefined to use instead a ratio of 
the vector vertex function for conserved and nonconserved currents. We show 
the results from implementing this method. 

We stress here that we are determining $Z_V$ very precisely and 
hence comparing values with uncertainties at the 0.1\% level. 
Previous work has compared values for $Z_V$ for nonconserved currents 
from methods that use 
Ward identities and the RI$'$-MOM 
scheme (for example~\cite{Becirevic:2004ny, Constantinou:2010gr}) 
and concluded that there was agreement at the 1\% level. 
Our more accurate results show clear disagreement, most 
obviously in the analysis for the conserved current.

Our earlier argument that 0.1\% accuracy is needed 
for renormalisation constants in pure lattice QCD can be 
extended when we study the impact of adding QED effects. 
When we allow the valence quarks to have  
electric charge (i.e. adding quenched QED to lattice QCD) 
we see a tiny impact (less than 0.1\%) on $Z_V$ using the HISQ action. 
We can now quantify and analyse this effect using the 
RI-SMOM scheme, having established that the nonperturbative $Z_V$ values behave 
correctly. 

The paper is laid out as follows: 
We first discuss in Section~\ref{sec:ward} the exact lattice vector 
Ward-Takahashi identity that gives the conserved vector current for the HISQ 
action. 
We then give a brief overview of the momentum-subtraction schemes, called RI-SMOM 
and RI$'$-MOM, that we will use (abbreviating the names to SMOM and MOM)
in Section~\ref{sec:MOM-schemes} and, following that, a brief description of our 
lattice set-up in Section~\ref{sec:latt}.
We show how the Ward-Takashashi identity works for the HISQ action 
in Section~\ref{subsec:wti-latt} so that the conserved current is 
not renormalised. This is then translated into the RI-SMOM scheme 
in Section~\ref{subsec:SMOMcons} where $Z_V=1$ is obtained for the 
conserved current at all $a$ and $\mu$ values. 
For RI$'$-MOM, however, condensate contributions are clearly evident in the 
$Z_V$ values for the conserved current  as shown in 
Section~\ref{subsec:MOMcons}. 
In Sections~\ref{subsec:SMOMloc} and~\ref{subsec:MOMloc} we 
demonstrate the impact of the protection from the Ward-Takahashi identity  
on the renormalisation factors for the simple 
local vector current ($\overline{\psi}\gamma_{\mu}\psi$ with the fields at the same 
space-time point). The difference between 
$Z_V^{\text{loc}}(\text{SMOM})$ 
and $Z_V^{\text{loc}}(\text{F(0)})$ is purely a discretisation effect; 
in the RI$'$-MOM scheme we demonstrate how to achieve 
the same outcome with a renormalisation factor that is a ratio between that 
for the local and conserved currents. 
In Section~\ref{sec:QED} we show the impact of quenched QED on 
the $Z_V$ values obtained for the local current in the RI-SMOM scheme 
and compare to our expectations based on the work in earlier sections. 
Finally, in Section~\ref{sec:conclusions} we discuss the implications of 
these results for ongoing and future calculations and give our conclusions.  
A similar picture to that for the local current 
is seen for the 1-link point-split vector current and we give the 
RI-SMOM results for this case
in~Appendix~\ref{app:1link}.  

We reiterate the shorthand notation that we will use 
for the renormalisation constants for clarity. 
$Z_V^{\text{x}}(\text{A})$ renormalises the
lattice vector current x (cons, loc, 1link) to  
match the continuum current (in e.g. $\overline{\text{MS}}$) 
and has been calculated in the scheme A (F(0), SMOM, MOM). 

\section{The vector Ward-Takahashi identity on the lattice} \label{sec:ward}

For both continuum and lattice~\cite{Karsten:1980wd, Bochicchio:1985xa} actions the derivation of the vector Ward-Takahashi identity proceeds from the observation that the path integral is
invariant under a local change of the fermion field variables $\psi$ and $\overline{\psi}$ 
(only) that has unit Jacobian. 
Then 
\begin{equation}
\label{eq:pathint}
\int \mathcal{D}\psi\mathcal{D}\overline{\psi}e^{-S[\psi^{\epsilon},\overline{\psi}^{\epsilon}]}f(\psi^{\epsilon},\overline{\psi}^{\epsilon})=\langle f(\psi,\overline{\psi}) \rangle \, .
\end{equation}
An example of such a transformation is to multiply $\psi$, say at point $x$, by
phase $e^{i\epsilon}$ and $\overline{\psi}(x)$ by $e^{-i\epsilon}$:
\begin{equation}
\label{eq:psieps}
\psi(z) \rightarrow \psi^{\epsilon}(z) \equiv \begin{cases}
e^{i\epsilon} \psi(x) & \mbox{for $z = x$} \\
\psi(x) & \mbox{for $z \ne x$} \, .
\end{cases}
\end{equation}
Expanding Eq.~(\ref{eq:pathint}) to first order in $\epsilon$ and 
denoting $\Delta X = X^{\epsilon}-X$ to this order gives 
\begin{equation}
\label{eq:deltapi}
\langle - \Delta S \cdot f + \Delta f \rangle = 0 \, .
\end{equation}

If we consider the path integral for the two point correlator 
$\langle \overline{\psi}(y_1) \psi(y_2) \rangle$ then 
$\Delta f$ becomes the difference of propagators from the 
points $y_1$ and $y_2$ to $x$. 
$\Delta S$ can be recast into the form $\Delta_{\mu}J^{\mu}$,
allowing us to identify the conserved current $J$ associated with $S$. 
We have
\begin{align} \label{eq:ward-pos} \langle \Delta_{\mu} J^{\mu}(x)
\overline{\psi}(y_1) \psi(y_2) \rangle &= \delta_{y_2,x} \langle
\overline{\psi}(y_1)\psi(x) \rangle \\ &- \delta_{y_1,x} \langle
\overline{\psi}(x)\psi(y_2) \rangle \nonumber . \end{align}
The right-hand side is zero unless $y_1$ or $y_2$ overlaps with $x$ (and 
not with each other). Note that $\Delta_{\mu} J^{\mu}(x)$ is centred on 
the point $x$.   

On the lattice $\Delta_{\mu}$ can be a simple forward ($\Delta_{\mu}^+$) 
or backward ($\Delta_{\mu}^-$) finite difference over one link. The current $J^{\mu}$
must then be chosen appropriately so that
\begin{eqnarray}
\label{eq:deltadef}
\Delta^{\mu}J_{\mu}(x) &\equiv& \Delta^{\mu, +}J_{\mu}^- = \sum_{\mu} \left(J_{\mu}^-(x+\hat{\mu}) - J_{\mu}^-(x)\right) \\
&\equiv& \Delta^{\mu, -}J_{\mu}^+ = \sum_{\mu} \left(J_{\mu}^+(x) - J_{\mu}^+(x-\hat{\mu})\right) \nonumber
\end{eqnarray}
We give $J_{\mu}^+$ for the HISQ action~\cite{Follana:2006rc} that we use in Appendix~\ref{app:cons-curr}.  
As discussed in the Introduction, it is rather complicated. 
It contains a number 
of 3-link terms because of the Naik term~\cite{Naik:1986bn} that removes tree-level $a^2$ errors 
in the action. The position-space 
Ward-Takahashi identity of Eq.~(\ref{eq:ward-pos}) provides a test 
of the implementation of the conserved current and we have 
checked that this works for our implementation 
exactly on a single gluon-field configuration 
for a variety of choices of $y_1$ and $y_2$.  

We can perform the exact Fourier transform on the lattice of
Eq.~(\ref{eq:ward-pos}). 
The left-hand side becomes 
\begin{eqnarray}
\label{eq:wardmomlhs}
&&(1-e^{iaq_{\mu}}) \times \\
&&\int d^4xd^4y_1 d^4y_2 e^{iqx}e^{-ip_1y_1}e^{ip_2y_2}\langle J^{\mu,+}(\tilde{x})\overline{\psi}(y_1)\psi(y_2) \rangle  \nonumber
\end{eqnarray}
where $a$ is the lattice spacing and we take $q=p_1-p_2$.
$\tilde{x}$ is the mid-point of the link between $x$ and $x+\hat{\mu}$.
The right-hand side becomes 
\begin{eqnarray}
\label{eq:wipropmom}
&&\int d^4x d^4y_1 e^{ip_1x}e^{-ip_1y_1} \langle \overline{\psi}(y_1)\psi(x) \rangle \nonumber \\
&&-\int d^4x d^4y_2 e^{-ip_2x}e^{ip_2y_2} \langle \overline{\psi}(x)\psi(y_2) \rangle \nonumber \\
&&\hspace{3em}\equiv S(p_1)-S(p_2) 
\end{eqnarray}
where $S$ is the quark propagator.
Then, multiplying on both sides by the product of inverse quark propagators 
we reach the lattice version of the 
standard expression for the Ward-Takahashi identity, 
\begin{equation} \label{eq:ward-mom}
 \frac{-2i}{a} \sin\left(\frac{aq_{\mu}}{2}\right) \Lambda_V^{\mu,+}(p_1,p_2) = -S^{-1}(p_1) + S^{-1}(p_2) .
\end{equation}
$\Lambda_V^{\mu,+}$ is the amputated vertex function for the 
vector current $J^{\mu,+}$ (absorbing a factor of $e^{iaq_{\mu}/2}$ into the 
vertex function since $J^{\mu,+}$ sits on a link). 
This equation is exact, 
gluon-field configuration by configuration, in lattice QCD
and we will demonstrate this for the HISQ action in Section~\ref{subsec:wti-latt}. 

As is well-known, Eq.~(\ref{eq:ward-mom}) tells us that any rescaling of the vertex by 
renormalisation on the left-hand side has to match rescaling of the inverse 
propagators on the right-hand side. This means that $J^{\mu,+}$ is not renormalised, i.e. 
that the renormalisation factor for this conserved current, 
$Z_V^{\text{cons}}$=1. Since this is also true for the 
conserved current in the continuum $\overline{\text{MS}}$ scheme 
then the matrix elements of the lattice conserved current will agree 
in the continuum limit with those in the $\overline{\text{MS}}$ scheme. 

A renormalised nonconserved vector current, written for example 
as $Z_V^{\text{loc}}V^{\text{loc},\mu}$ for a local 
current, obeys the same 
equations as $J^{\mu,+}$ since it is by definition the same operator 
up to discretisation effects on the lattice~\cite{Vladikas:2011bp}. 
For the HISQ action 
\begin{equation}
\label{eq:cons-loc}
J_{\mu}^+  = Z_V^{\text{loc}}V^{\text{loc}}_{\mu} + \mathcal{O}(a^2) .
\end{equation}
Again this is well-known, but we point it out here because it has 
implications for the accuracy of the determination of 
$Z_V^{\text{loc}}$ on the lattice. It means that, if $Z_V^{\text{loc}}$ 
is determined by a procedure which uses the Ward-Takahashi identity 
and gives 1 for the 
renormalisation of $J^{\mu,\pm}$, then 
$Z_V^{\text{loc}}$ must be
free of systematic errors 
from nonperturbative (condensate) 
artefacts in the continuum limit because these must cancel between 
the left- and righthand sides of Eq.~\ref{eq:ward-mom}. 
$Z_V^{\text{loc}}$ can in principle 
be determined by substituting $Z_V^{\text{loc}}V^{\text{loc}}$ into 
the lefthand side of Eq.~(\ref{eq:ward-mom}) for any $p_1$ and $p_2$. 
Hadronic matrix elements of $Z_V^{\text{loc}}V^{\text{loc}}$ will then 
differ from the results in the continuum purely by discretisation 
effects (which will depend on $p_1$ and $p_2$) that can be extrapolated 
away straightforwardly using results at multiple values of the lattice spacing. 
The $Z_V$ so obtained is completely nonperturbative. 

Using Eq.~(\ref{eq:ward-mom}) in its full generality is unnecessarily 
complicated and there are lattice QCD methods that make use of it in 
specific, and simpler, kinematic configurations. 

As $q \rightarrow 0$ the identity of Eq.~(\ref{eq:ward-mom}) can be used to show 
that the vector form factor for the conserved current 
between quark or hadron states of the same momentum 
will be unity.  The inverse of the vector form factor at the same 
kinematic point for a nonconserved 
current then gives its $Z_V$ value.  This method clearly satisfies 
the criteria above for an exact determination of $Z_V$. 

We now discuss momentum-subtraction renormalisation schemes 
on the lattice and the extent to which they make use of Eq.~(\ref{eq:ward-mom}). 

\section{Momentum-subtraction schemes used on the lattice} \label{sec:MOM-schemes}

Momentum-subtraction schemes are useful intermediate schemes between 
the lattice regularisation and the continuum $\overline{\text{MS}}$ scheme 
in which it is now standard to quote results for scheme-dependent quantities. 
If the same momentum-subtraction scheme is implemented both in lattice QCD and in 
continuum QCD then the continuum limit of the lattice results will be in 
the continuum momentum-subtraction scheme (and should be independent of lattice 
action at that point). They can then be converted to 
the $\overline{\text{MS}}$ scheme using continuum QCD perturbation theory. 

A momentum-subtraction scheme imposes renormalisation conditions on matrix elements 
between (in the cases we consider) external quark states so that the tree-level 
result is obtained, i.e.\ $Z_{\Gamma}$ is defined by 
\begin{equation}
\label{eq:momgen}
Z_{\Gamma} \langle p_1 | O_{\Gamma} | p_2 \rangle = \langle p_1 | O_{\Gamma} | p_2 \rangle_{\text{tree}}
\end{equation}
for some operator $O_{\Gamma} = \overline{\psi} \Gamma \psi$, and $\langle p_1 |$ and 
$|p_2 \rangle$ external quark states with momenta $p_1$ and $p_2$, typically taken to have 
large magnitude. This calculation must of course be done in a fixed gauge, and this 
is usually taken to be Landau gauge,  
which can be straightforwardly implemented in lattice QCD. 
Effects from 
the existence of Gribov copies under the gauge-fixing could arise in general; 
here we show that there are no such effects
for $Z_V$ determined using the Ward-Takashahi 
identity. 

Here we will concentrate on the RI-SMOM scheme~\cite{Aoki:2007xm,Sturm:2009kb}. 
This scheme uses a symmetric kinematic configuration with 
only one scale so that $p_1^2=p_2^2=q^2=\mu^2$ (where $q=p_1-p_2$). 
The wavefunction renormalisation is defined (using continuum notation) by 
\begin{equation}
\label{eq:zqdef}
Z_q = - \frac{i}{12p^2}\mathrm{Tr}(\slashed{p}S^{-1}(p)) .
\end{equation}
The vector current renormalisation follows from requiring
\begin{equation}
\label{eq:zvdef}
  \frac{Z_V}{Z_q}\frac{1}{12q^2} \mathrm{Tr}(q_{\mu}\Lambda_V^{\mu}(p_1,p_2)\slashed{q}) = 1 . 
\end{equation}
The traces here are over spin and colour and normalisations are chosen so 
that $Z_q=Z_V=1$ at tree-level. 
The equations above are given for the continuum SMOM scheme. 
On the lattice we must take care to define the appropriate discretisation
for $q_{\mu}$ and $q^2$ in the various places that they appear. 
Below we will see what form $q_{\mu}$ must take in combination 
with the vertex function for the conserved current. 

The RI-SMOM scheme was defined with the vector Ward-Takahashi identity in 
mind~\cite{Sturm:2009kb}. This reference shows how the identity defines the projectors 
needed for the vector vertex function in the continuum (given in in Eq.~(\ref{eq:zvdef})) so 
that $Z_V=1$ for the conserved current. Here we repeat this exercise, but 
now on the lattice. Returning to the Ward-Takahashi identity in Eq.~(\ref{eq:ward-mom})
we can multiply both sides by $\slashed{\hat{q}}$ and take the trace whilst dividing 
by $\hat{q}^2$ (with $\hat{q}$ a discretisation of $q$ to be defined later). 
This gives
\begin{align} \label{eq:rismom-der}
  &\frac{1}{12\hat{q}^2}\frac{-2i}{a}\sin(aq_{\mu}/2) \mathrm{Tr}(\Lambda_V^{\mu,+}\slashed{\hat{q}})\\ &=  \frac{1}{12\hat{q}^2} [-\mathrm{Tr}(S^{-1}(p_1)\slashed{\hat{q}}) + \mathrm{Tr}(S^{-1}(p_2)\slashed{\hat{q}})] \nonumber .
\end{align}
We can simplify the right-hand side assuming that the inverse propagator takes  
the general form $S^{-1}(p) = i\slashed{p}\Sigma_V(p^2) + \Sigma_S(p^2)$ in the continuum 
(from relativistic invariance).
Then, 
for the SMOM kinematics,   
\begin{equation} \label{eq:rismom-req}
  \mathrm{Tr}(S^{-1}(p_1)\slashed{q}) - \mathrm{Tr}(S^{-1}(p_2)\slashed{q}) = \mathrm{Tr}(S^{-1}(q)\slashed{q}) .
\end{equation}
On the lattice this formula could be broken by discretisation effects. We do not 
see noticeable effects of this kind with the HISQ action, however, as we will discuss 
in Section~\ref{subsec:SMOMcons}. 

Using Eq.~(\ref{eq:rismom-req}) in Eq.~(\ref{eq:rismom-der}) 
and multiplying by $i$ 
then gives, 
from the Ward-Takahashi identity 
\begin{equation} \label{eq:rismom-zv1}
  \frac{1}{12\hat{q}^2}\frac{2}{a}\sin(aq_{\mu}/2) \mathrm{Tr}(\Lambda_V^{\mu,+}\slashed{\hat{q}}) =  -\frac{i}{12\hat{q}^2} \mathrm{Tr}(S^{-1}(q)\slashed{\hat{q}}) \, .
\end{equation}
From Eq.~(\ref{eq:zqdef}) we see that the righthand-side of this expression is 
$Z_q$ in the RI-SMOM scheme. 
Comparing the left-hand side 
to Eq.~(\ref{eq:zvdef}) we see that this is $Z_q/Z_V^{\text{cons}}$ 
in the RI-SMOM scheme
where $Z_V^{\text{cons}}$ is the $Z_V$ factor for the conserved current
and the Ward-Takahashi identity requires us to discretise $q_{\mu}$    
as $2\sin(aq_{\mu}/2)/a$ ($\hat{q}$ is defined in Eq.~(\ref{eq:qhatdef})). 
Then, from Eq.~(\ref{eq:rismom-zv1}), we expect that $Z_V^{\text{cons}}(\text{SMOM})=1$ 
on the lattice and no 
further renormalisation is needed to match to $\overline{\text{MS}}$. 
Notice that this works for any value of $q$. 

We will show by explicit calculation that $Z_V^{\text{cons}}(\text{SMOM})=1$ 
for the HISQ action
in Section~\ref{subsec:SMOMcons}. This is not true configuration by 
configuration, however. It does require an average over gluon fields. 

Another popular momentum-subtraction scheme is 
RI$'$-MOM~\cite{Martinelli:1994ty, Chetyrkin:1999pq}, abbreviated here to MOM. 
In this scheme $Z_q$ is defined in the same way, by Eq.~(\ref{eq:zqdef}), but 
$Z_V$ is defined by a different projector for the vector vertex function and the kinematic 
configuration for the MOM case is $p_1=p_2=p$ so that $q=0$. 
Instead of Eq.~(\ref{eq:zvdef}) we have, in the MOM scheme, 
\begin{equation}
\label{eq:zvmomdef}
  \frac{Z_V}{Z_q}\frac{1}{12} \mathrm{Tr}(\gamma_{\mu}\Lambda_V^{\mu}(p)) = 1 . 
\end{equation}
Since this scheme does not correspond to a Ward-Takahashi identity, $Z_V$ determined 
this way needs further renormalisation to match to the $\overline{\text{MS}}$ 
scheme. More problematically,  
as we will show in 
Section~\ref{sec:latt}, $Z_V^{\text{cons}}(\text{MOM})$ for the HISQ action is 
significantly different from 1 and is contaminated by nonperturbative condensate effects. 

The RI-SMOM$_{\gamma_{\mu}}$ scheme~\cite{Sturm:2009kb} is similar to RI$'$-MOM above but uses the 
SMOM kinematics with $p_1^2=p_2^2=q^2$. 

To calculate the renormalisation constants for nonconserved currents we 
must combine the calculation of the vector vertex function for that current 
(Eq.~(\ref{eq:zvdef}) and appropriate modifications of it as described in the text)
with the calculation of the wave-function renormalisation (Eq.~(\ref{eq:zqdef})). 
We describe the results for the HISQ local vector current in the SMOM scheme 
in Section~\ref{subsec:SMOMloc}. 
We are able to show that the renormalisation factor for the local 
vector current in the SMOM scheme differs from that using the form factor method purely 
by discretisation effects, demonstrating that it is an exact form of $Z_V$. 
The discretisation effects depend on $q$ but the method is exact for any $q$;
this is in contrast to the usual idea of a `window' of $q$ values to be used 
in momentum-subtraction schemes on the lattice~\cite{Martinelli:1994ty}.

The RI$'$-MOM scheme is not exact, as discussed above. 
We show in Section~\ref{subsec:MOMloc} that a modification of the method (reverting 
to one of the original suggestions in~\cite{Martinelli:1994ty}) does, 
however, give an exact
$Z_V$. 

There are technical issues associated with implementing momentum-subtraction schemes for 
staggered quarks that we will not discuss here. We use the techniques 
developed in~\cite{Lytle:2013qoa} and summarised again in~\cite{Lytle:2018evc} in the 
context of the RI-SMOM scheme. We will only discuss here specific issues that arise 
in the context of the vector current renormalisation.  

\section{The Lattice QCD calculation}
\label{sec:latt}

We perform calculations on $n_f=2+1+1$ gluon field configurations generated 
by the MILC collaboration~\cite{Bazavov:2010ru,Bazavov:2012xda} listed in 
Table~\ref{tab:ensembles}. These ensembles
use an improved gluon action which removes discretisation errors 
through $\mathcal{O}(\alpha_sa^2)$~\cite{Hart:2008sq}. They include the effect of 
$u/d$, $s$ and $c$ quarks in the sea using the HISQ action~\cite{Follana:2006rc}. 

All gauge field configurations used are numerically fixed to Landau 
gauge by maximising the trace over the gluon field link with a gauge
fixing tolerance of $\epsilon=10^{-14}$. This is enough to remove the 
difficulties related to loose gauge fixing discussed 
in~\cite{Lytle:2018evc}. 

We use broadly the same
calculational set up as in \cite{Lytle:2018evc} but here we are considering vector current 
vertex functions rather than scalar ones. 
To implement momentum-subtraction schemes for staggered quarks we need 
to use momenta within a reduced Brillouin zone~\cite{Lytle:2013qoa} 
\begin{equation}
\label{eq:redbrill}
-\pi/2 \le ap \le \pi/2 .
\end{equation}
For each momentum $ap_1$ or $ap_2$ we then calculate
propagators or vertex functions with 16 copies 
of that momentum, $ap_1+\pi A$ and $ap_2 +\pi B$ where 
$A$ and $B$ are four-vectors 
composed of 0s and 1s. This then enables us to do the 
traces over spin for specific `tastes' of vector current 
implied by equations such as 
Eq.~(\ref{eq:rismom-zv1}). There is also a trace over colour 
in this equation so the $S^{-1}(q)$ factor on the righthand side, for example, 
is actually a $48\times48$ matrix.  
Where necessary we will use the notation of~\cite{Lytle:2013qoa} 
to denote specific spin-tastes. As an example 
$\overline{\overline{\gamma_{\mu} \otimes I}}$ is the 
$16 \times 16$ matrix of 0s and 1s that projects onto 
a taste-singlet vector in $AB$ space. 

\begin{table}
\caption{Simulation parameters for the MILC gluon field
ensembles that we use, labelled by set number in the first column.
$\beta=10/g^2$ is the bare QCD coupling and $w0/a$ gives 
the lattice spacing~\cite{Borsanyi:2012zs}, 
using $w_0$ = 0.1715(9) fm~\cite{fkpi} determined from 
the $\pi$ meson decay constant, $f_{\pi}$. 
Note that, for each group of ensembles at a given 
value of $\beta$ we use the $w_0/a$ value corresponding 
to the physical sea quark mass limit~\cite{Lytle:2018evc}, 
using results from~\cite{Chakraborty:2014aca}. 
$L_s$ and $L_t$
give the lattice dimensions. $am_l^{\sea}$, $am_s^{\sea}$
and $am_c^{\sea}$ give the sea quark masses in lattice units.
Set 1 will be referred to in the text 
as `very coarse', sets 2--5 as `coarse', set 6 as `fine'
and set 7 as `superfine'. 
}
\label{tab:ensembles}
\begin{ruledtabular}
\begin{tabular}{lllllllll}
Set & $\beta$ & $w_0/a$ & $L_s$ & $L_t$ & $am_l^{\sea}$ & $am_s^{\sea}$ & $am_c^{\sea}$ \\
\hline
1 & 5.80 & 1.1322(14) & 24 & 48 & 0.00640 & 0.0640 & 0.828 \\
2 & 6.00 & 1.4075(18) & 24 & 64 & 0.0102 & 0.0509 & 0.635 \\
3 & 6.00 & 1.4075(18) & 24 & 64 & 0.00507 & 0.0507 & 0.628 \\
4 & 6.00 & 1.4075(18) & 32 & 64 & 0.00507 & 0.0507 & 0.628  \\
5 & 6.00 & 1.4075(18) & 40 & 64 & 0.00507 & 0.0507 & 0.628  \\
6 & 6.30 & 1.9500(21) & 48 & 96 & 0.00363 & 0.0363 & 0.430 \\
7 & 6.72 & 2.994(10) & 48 & 144 & 0.0048 & 0.024 & 0.286  \\
\end{tabular}
\end{ruledtabular}
\end{table}

Twisted boundary conditions are utilised to 
give the incoming and outgoing quarks arbitrary
momenta~\cite{twist, Arthur:2010ht}. For the SMOM 
kinematics we take, with ordering $(x,y,z,t)$, 
\begin{eqnarray} 
\label{eq:SMOMmom}
ap_1 &=& (a\mu, 0, a\mu, 0)/\sqrt{2} \\
ap_2 &=& (a\mu, -a\mu, 0, 0)/\sqrt{2} \, .\nonumber 
\end{eqnarray} 
For the MOM kinematics we take $ap_2=ap_1$. 
A range of $a\mu$ values are chosen at each lattice spacing, 
satisfying Eq.~(\ref{eq:redbrill}). This allows us to 
reach $\mu$ values of 3 GeV on coarse lattices and 4 GeV on 
fine and superfine lattices~\cite{Lytle:2018evc}. 
The $\mu$ values can be tuned very accurately (to 3 dec. places). 

Relatively small samples (20 configurations) give small statistical 
uncertainties for $Z_V$ at the $\mu$ values that we use (with momentum 
sources for the propagators). 
A bootstrap method is used to 
estimate all uncertainties and include correlations between results at 
different $\mu$ values on a given ensemble. Bootstrap samples are formed for
each $Z_q$ and each ${\Lambda_V}$ and the bootstrap averages are then fed into the  
ratio to determine $Z_V$. 

All of our results are determined at small but non-zero valence quark mass. 
Degenerate masses are used for the incoming and outgoing quarks
(but note that there is no need for the calculation of disconnected contributions). 
As the momentum-subtraction schemes that we consider are in principle defined at 
zero valence quark mass (but direct calculation at this point will have 
finite-volume issues)
it is necessary to calculate each $Z_V$ at different quark 
masses and then extrapolate to the $am_{\mathrm{val}}=0$ point.
To do this we perform all calculations at three masses 
corresponding to the light sea quark mass on a given ensemble, $am_l$, and at
$2am_l$ and $3am_l$. 
Dependence on $am_{\mathrm{val}}$ can come from 
discretisation effects and from the contribution of nonperturbative 
condensate terms.

We follow the procedure used for $Z_m$ in~\cite{Lytle:2018evc} 
and extrapolate $Z_V$ results
using a polynomial in $am_{\mathrm{val}}/am_s$:
\begin{align}
\label{eq:massextrap}
  Z_V(am_{\mathrm{val}},\mu) = Z_V(\mu) &+ d_1(\mu)\frac{am_{\mathrm{val}}}{am_s} \\ &+ d_2(\mu) \left( \frac{am_{\mathrm{val}}}{am_s} \right)^2 .\nonumber
\end{align}
We find no need for higher powers of $am_{\mathrm{val}}/am_s$ here as the valence 
mass dependence of $Z_V$ is observed to be very mild
in all cases. 
For the priors for the coefficients $d_i$ we use $\{0\pm0.1,0\pm0.01 \}$ at $\mu=2$
GeV with the widths decreased according to $\mu^{-2}$.

Any sea quark mass dependence should be suppressed relative to the 
valence mass dependence by powers of $\alpha_s$ and this was observed in~\cite{Lytle:2018evc}. 
As the valence mass
dependence is already negligible the sea mass dependence should be tiny 
here and we ignore it. 

\subsection{The Ward-Takahashi identity on the lattice}
\label{subsec:wti-latt}

\begin{figure}[ht]
  \includegraphics[width=0.47\textwidth]{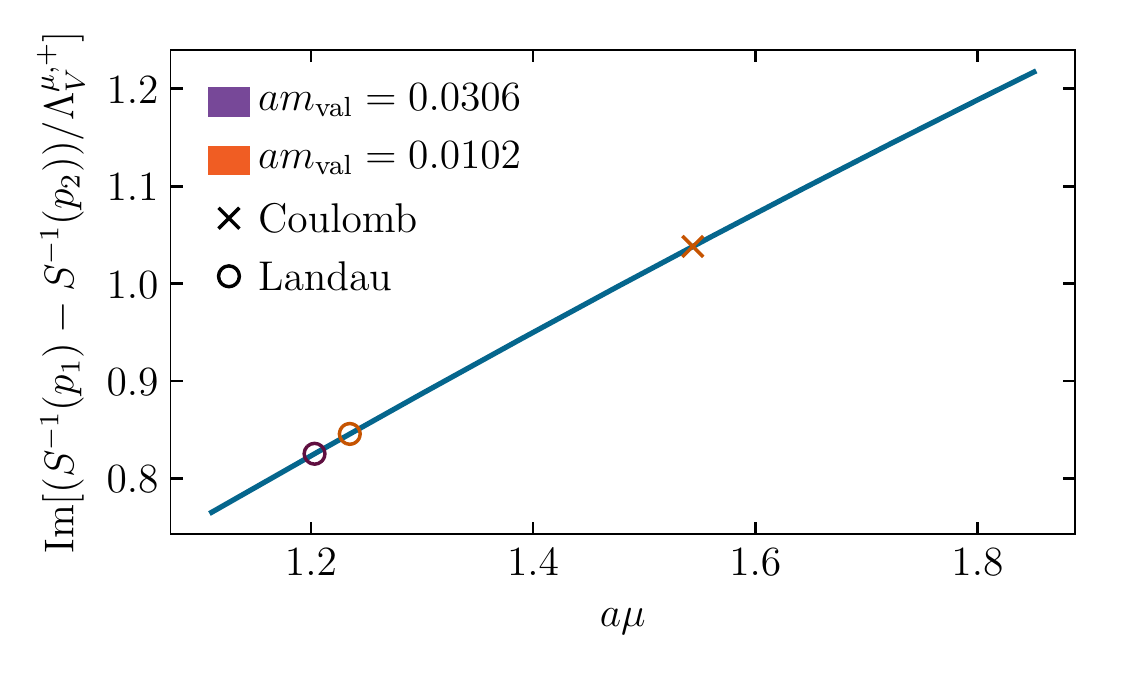}
  \caption{
Demonstration of the vector Ward-Takahashi identity in 
momentum space (Eq.~\ref{eq:ward-mom}) 
for HISQ quarks on the lattice on a single gluon field configuration from Set 2. 
The plot shows the ratio
of the righthand side of this equation 
to the amputated vertex function for the conserved vector current on the lefthand side 
for the SMOM kinematic configuration, Eq.~(\ref{eq:SMOMmom}).  
This is a matrix equation and this plots shows the result of averaging 
over all matrix components (which agree) and the two non-zero components 
of $q_{\mu}$.
The solid line is the value of $2\sin(aq_{\mu}/2)$
 for a non-zero component of $q_{\mu}$. 
  The points correspond to
  lattice results for the ratio on a single configuration with crosses
  giving Coulomb gauge-fixed results and the circles Landau gauge-fixed results. 
Orange points correspond to a valence mass of $am_{\mathrm{val}}=0.0306$
  while purple points correspond to 0.0102. 
The Ward-Takahashi identity requires these points to lie on the line as they do.}
  \label{fig:ward-id}
\end{figure}

In this Section we test the exact lattice Ward-Takahashi identity for 
HISQ quarks, i.e. Eq.~(\ref{eq:ward-mom}).  If we have correctly implemented 
the lattice conserved vector current, this equation is true as a $3 \times 3$ matrix 
in colour space. It is also true for any $p_1$ and $p_2$ 
(except that it reduces to 0=0 for $p_1=p_2$), any values of the quark mass and any gauge. 
We test it for the SMOM kinematic configuration of Eq.~(\ref{eq:SMOMmom}).  

Figure~\ref{fig:ward-id} shows the results as a ratio of the difference of inverse 
propagators on the righthand side of Eq.~(\ref{eq:ward-mom}) to the 
amputated vertex function for the conserved vector current on the lefthand side. 
This is averaged over colour components (which all agree) and summed over 
the two non-zero components of $q_{\mu}$ (which take the same value 
$a\mu/\sqrt{2}$ in each of the $y$ and $z$ directions for the SMOM  kinematics). 
The Ward-Takahashi identity (Eq.~(\ref{eq:ward-mom})) requires this ratio 
to be exactly equal to $2\sin[a{\mu}/(2\sqrt{2})]$, which is plotted as the line. 

The plot shows that this expectation works to high precision 
(double precision accuracy here), on a single
configuration taken as an example from Set 2. 
Results are given for three different $a\mu$ values with 
two different valence quark masses 
and in two different gauges. The agreement between the points and 
the line demonstrates the Ward-Takashi identity 
working explicitly on the lattice for the conserved HISQ current 
of Eq.~(\ref{eq:Jcons}). 
The agreement seen in two different gauges is evidence that 
the Ward-Takahashi identity works in any gauge, as it must, and therefore 
its operation is also independent of any Gribov copy issue in the gauge-fixing 
procedure. 

\subsection{$Z_V$ for the conserved current in the RI-SMOM scheme}
\label{subsec:SMOMcons}

\begin{figure}[ht]
  \includegraphics[width=0.47\textwidth]{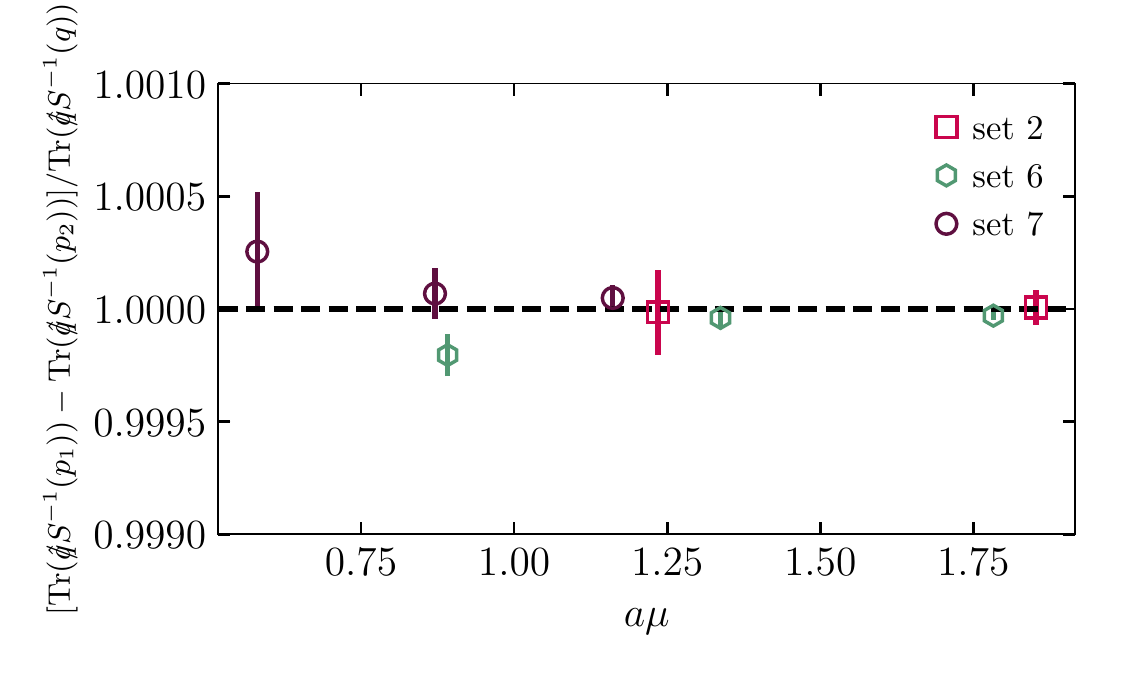}
  \caption{ A test of the expression for the difference of inverse propagators with 
momentum $p_1$ and $p_2$ in Eq.~(\ref{eq:rismom-req}). We show results on 
coarse, fine and superfine lattices (sets 2, 5 and 7) for a variety of $\mu$ values in lattice 
units where $|ap_1|=|ap_2|=|aq|=a\mu$. 
}
  \label{fig:prop-wi-test}
\end{figure}

\begin{figure}[ht]
  \includegraphics[width=0.47\textwidth]{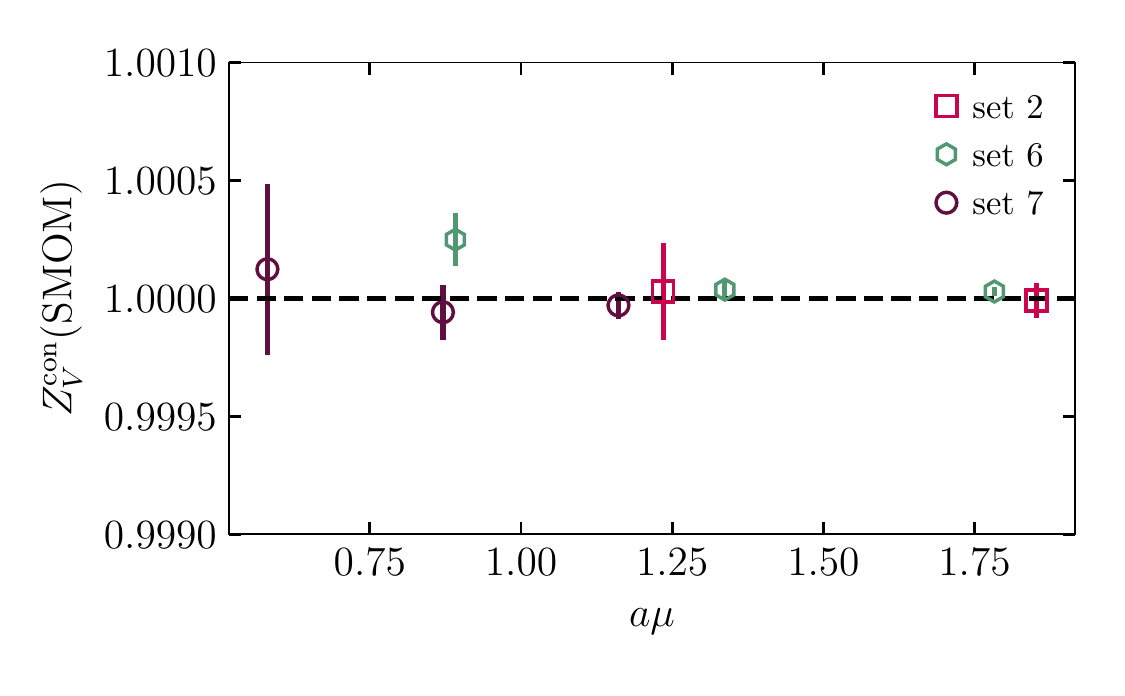}
  \caption{ The $Z_V$ value obtained for the conserved vector current 
in the RI-SMOM scheme on coarse, fine and superfine gluon field configurations 
(sets 2, 5 and 7). 
Values are given for a variety of $\mu$ values in lattice 
units where $|ap_1|=|ap_2|=|aq|=a\mu$. 
}
  \label{fig:Zv-con-smom}
\end{figure}

To determine $Z_V$ for the HISQ conserved current 
in the RI-SMOM scheme we adapt Eqs~(\ref{eq:zqdef}) and~(\ref{eq:zvdef}) 
to the case of staggered quarks on the lattice, as partly discussed already
in Section~\ref{sec:MOM-schemes}. 
For staggered quarks the inverse propagator is a taste-singlet~\cite{Lytle:2013qoa} 
and so the HISQ version of Eq.~(\ref{eq:zqdef}) is 
\begin{equation}
\label{eq:zqdefhisq}
Z_q(q) = -\frac{i}{48} \sum_{\mu} \frac{a\hat{q}_{\mu}}{(a\hat{q})^2}\Tr\left[ \overline{\overline{(\gamma_{\mu}\otimes I)}}S^{-1}(q)\right] . 
\end{equation}
The trace is now over colour and $AB$-space index. 
$\hat{q}$ is given by 
\begin{equation}
\label{eq:qhatdef}
a\hat{q}_{\mu} = \sin(aq_{\mu}) + \frac{1}{6}\sin^3(aq_{\mu}) .
\end{equation}
This choice is dictated by the momentum-subtraction 
requirement that $Z_q$ should be 1 in 
the non-interacting (tree-level) case and the fact that the derivatives in the HISQ 
action are improved through $\mathcal{O}(a^2)$~\cite{Follana:2006rc}. 
Likewise the HISQ calculation for $Z_V$ for this case is given by 
\begin{equation}
\label{eq:zvdefhisq}
\frac{Z_q(q)}{Z_V(q)} = \frac{1}{48} \sum_{\mu,\nu} 2\sin(aq_{\mu}/2)\frac{a\hat{q}_{\nu}}{(a\hat{q})^2}\Tr\left[ \overline{\overline{(\gamma_{\nu}\otimes I)}}\Lambda^{\mu,+}_V\right] . 
\end{equation}

In Sec.~\ref{sec:MOM-schemes} it was shown how the 
Ward-Takahashi identity leads to the exact expression of 
Eq.~(\ref{eq:rismom-der}) on the lattice when the conserved 
current is used in the vertex function. 
In order to obtain $Z_V=1$ for the conserved current we 
also need Eq.~(\ref{eq:rismom-req}) to be satisfied exactly. 
In Fig.~\ref{fig:prop-wi-test} we give a test of this 
relationship. The figure shows the ratio of the difference of 
the two inverse propagators with momentum $p_1$ and $p_2$ 
to that of the propagator with momentum $q$, where the inverse 
propagators are multiplied by $\slashed{\hat{q}}$ and the 
trace taken. We use $\hat{q}$ here (Eq.~(\ref{eq:qhatdef})) instead of simply $q$ 
to be consistent with what we use in the determination of $Z_q$ in 
Eq.~(\ref{eq:zqdefhisq}) above. The results for the ratio plotted would 
be the same for $q$ as for $\hat{q}$.  
The results for the ratio in Fig.~\ref{fig:prop-wi-test} are seen 
to be consistent with 1.0 to better than 0.05\%. The statistical 
uncertainties plotted are from a bootstrap over results from 
20 gluon field configurations.  

Figure~\ref{fig:prop-wi-test} shows that discretisation effects in 
the HISQ action have no effect on Eq.~(\ref{eq:rismom-req}) at 
the level of accuracy to which we are working. 
There are no tree-level $a^2$ errors 
with HISQ~\cite{Follana:2006rc} and there is a U(1) axial symmetry; 
both of these constrain 
the form that discretisation effects can take~\cite{Lytle:2013qoa}. 
A further constraint comes from the 
form for the $p_1$ and $p_2$ momenta (and $q$) 
to achieve the SMOM kinematics. Each has only two 
non-zero momentum components, as shown in Eq.~(\ref{eq:SMOMmom}). 
This means, for example, that discretisation errors in 
$S^{-1}$ containing 3 different $\gamma$ matrices 
and associated momenta are zero. 

Figure~\ref{fig:Zv-con-smom} shows the resulting $Z_V$ value 
obtained for the conserved vector current in the RI-SMOM 
scheme, combining the results from Eqs~(\ref{eq:zvdefhisq}) 
and~(\ref{eq:zqdefhisq}) and performing the extrapolation to 
zero quark mass as described in Sec.~\ref{sec:latt} (this has very little 
impact). 
The value obtained for $Z_V$ for the conserved current 
is 1 to better than 0.05\% at all $\mu$ values.  
Fitting the results shown in Fig.~\ref{fig:Zv-con-smom} to a constant 
value of 1.0 returns a $\chi^2/\text{dof}$ of 1.3 for 8 degrees of 
freedom ($Q=0.26$).

\subsection{$Z_V$ for the conserved current in the RI$'$-MOM scheme}
\label{subsec:MOMcons}

\begin{table}
\caption{Conversion factors from the continuum RI$'$-MOM scheme to $\overline{\mathrm{MS}}$
at the $\mu$ values used in this calculation, 
calculated with $n_{\mathrm{f}}=4$ using the results of~\cite{Chetyrkin:1999pq}.
Results for $Z_V$ obtained on the lattice with the standard RI$'$-MOM approach 
must be multiplied by these values to give results in the $\overline{\mathrm{MS}}$ scheme 
in the continuum limit.}  
\label{tab:RI-MSbar-conv}
\begin{tabular}{ll}
$\mu$ [GeV] & $Z_V^{\overline{\mathrm{MS}}/\mathrm{RI}'\mhyphen \mathrm{MOM}}$ \\
\hline
2 & 0.99118(38) \\
2.5 & 0.99308(26) \\
3 & 0.99420(20) \\
4 & 0.99549(14) \\
\end{tabular}
\end{table}

\begin{figure}
  \includegraphics[width=0.47\textwidth]{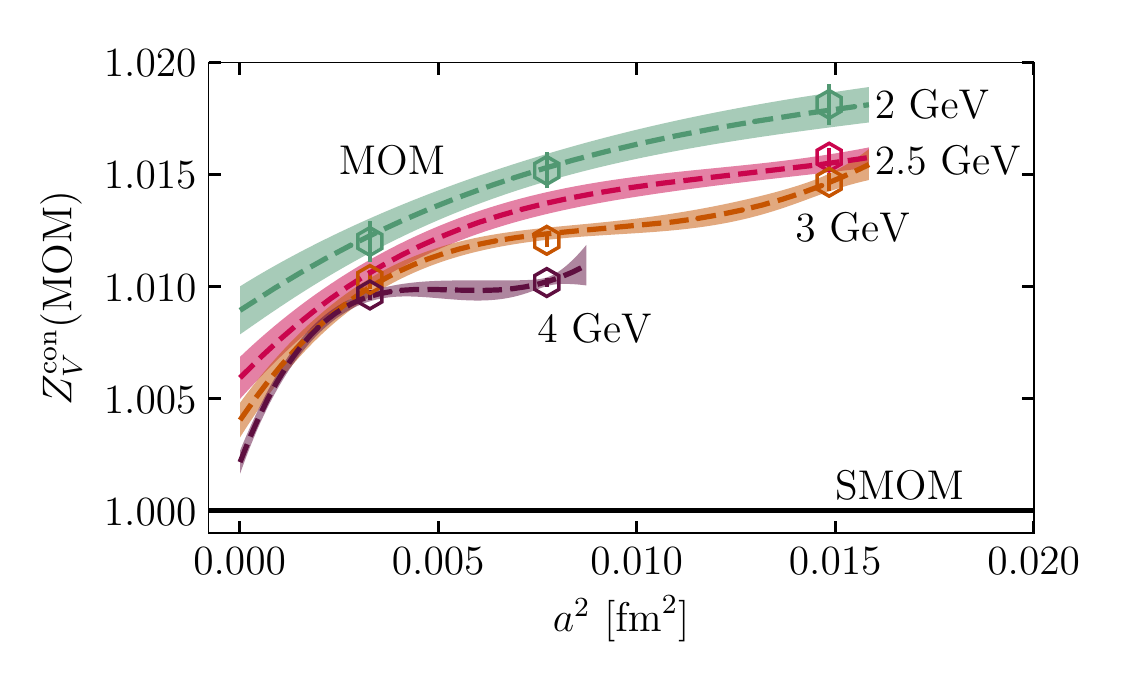}
  \caption{Points labelled `MOM' show 
the renormalisation factor for the HISQ conserved vector current which takes 
results from the lattice scheme to the $\overline{\text{MS}}$ scheme obtained
using a lattice calculation in the RI$'$-MOM scheme. 
These should be contrasted with results obtained using a lattice calculation in 
the RI-SMOM scheme which give value 1.0, shown as the black line labelled `SMOM'. 
Points are given for $\mu$ values 2, 2.5, 3 and 4 GeV, indicated by different colours, 
and on coarse, fine and superfine lattices. 
The fit shown (see Eq.~(\ref{eq:MOM-con-fit})) accounts for
  discretisation errors and condensate contributions, which prove to be necessary for a good fit. 
The separation of the results for different $\mu$ values in the continuum limit (at $a=0$) 
is a result of the condensate contributions that appear in $Z_V$ 
when calculated in the RI$'$-MOM scheme on the lattice. 
}
  \label{fig:MOM-con}
\end{figure}

We now turn to the renormalisation of the conserved current in 
the standard RI$'$-MOM scheme where a very different picture emerges. 

The kinematic conditions in the MOM scheme are that the incoming 
and outgoing quark fields for the vertex function should have 
the same momentum, $ap_1=ap_2$ so that $aq=0$. We will denote this 
momentum by $ap$ with $|ap|=a\mu$. We take the form of $ap$ to be that 
of $ap_1$ in the SMOM scheme (Eq.~(\ref{eq:SMOMmom})).  
To implement the RI$'$-MOM scheme we determine the wave-function 
renormalisation, $Z_q(p)$, in the same way as for the RI-SMOM scheme 
using Eq.~(\ref{eq:zqdefhisq}). 
To determine $Z_V$ we use
\begin{equation}
\label{eq:zvmomdefhisq}
\frac{Z_q}{Z_V} = \frac{1}{48} \frac{1}{V^{\text{cons}}_{\gamma\otimes I}}\sum_{\mu} \Tr\left[ \overline{\overline{(\gamma_{\mu}\otimes I)}}\Lambda^{\mu,+}_V\right] . 
\end{equation}
This uses the RI$'$-MOM vector vertex projector, 
which is simply $\gamma_{\mu}$ (see Eq.~(\ref{eq:zvmomdef})), expressed 
here in the appropriate taste-singlet form for implementation with staggered quarks. 
Because the conserved current is a point-split operator 
the tree-level vertex function is not simply 1. We therefore need to divide 
by the tree-level matrix element for the conserved current 
that we denote here $V^{\text{cons}}_{\gamma_{\mu} \otimes I}$. 
How to calculate these tree-level factors is discussed in~\cite{Lytle:2013qoa}. 
We have
\begin{equation}
\label{eq:constree}
V^{\text{cons}}_{\gamma \otimes I}= 
\prod_{\mu} \left ( \frac{9}{8} \mathrm{cos}(ap_{\mu} (S-T)_{\mu}) 
 - \frac{1}{8} \mathrm{cos}(3ap_{\mu} (S-T)_{\mu}) \right) .
\end{equation}
The spin-taste 4-vector $S-T$ is composed of 1s and 0s. For the taste-singlet 
vector it takes 
value 1 for component $\mu$ and 0 otherwise. So the only components 
of the product that do not take value 1 are those for component $\mu$ that 
matches the direction of the current, provided that $ap$ has a non-zero 
component in that direction.   

Because the RI$'$-MOM scheme is not based on the Ward-Takahashi identity 
$Z_V$ will not be 1 for the conserved current. This means that to reach 
the $\overline{\text{MS}}$ scheme, even for the continuum RI$'$-MOM scheme, requires an 
additional renormalisation factor. The renormalisation factor that takes 
the lattice vector current to the continuum is then  
\begin{equation}
\label{eq:msbarmom}
Z_V(\mathrm{MOM}) = Z_V^{\overline{\text{MS}}/{\text{RI}'\text{-MOM}}}Z_V^{\text{MOM,raw}} .
\end{equation}
$Z_V^{\text{MOM,raw}}$ is the raw renormalisation factor calculated using 
Eq.~(\ref{eq:zvmomdefhisq}) on the lattice. 
The factor $Z_V^{\overline{\text{MS}}/\text{MOM}}$ can be determined 
from the 
perturbative QCD expansions in the continuum 
for the conversion between RI$'$-MOM and 
RI-MOM given in~\cite{Chetyrkin:1999pq} (see~\cite{Huber:2010zza} and 
the Appendix of~\cite{Lytle:2013qoa}). 
The values needed for our $\mu$ values are given in 
Table~\ref{tab:RI-MSbar-conv}; 
they are all close to 1 since the expansion starts at $\mathcal{O}(\alpha_s^2)$. 

Figure~\ref{fig:MOM-con} shows our results for $Z_V$ for the conserved 
HISQ current obtained by implementing the RI$'$-MOM scheme on the 
lattice. We have converted the $Z_V$ to the value that takes the lattice 
results to the $\overline{\text{MS}}$ scheme 
using Eq.~(\ref{eq:msbarmom}). 
Results are shown, after extrapolation to zero valence quark mass, 
at a variety of $\mu$ values from 2 GeV to 4 GeV and 
at three different values of the lattice spacing. 
It is immediately clear that the values of $Z_V^{\text{cons}}(\text{MOM})$ are not 1. 
This is in sharp contrast to results in the RI-SMOM scheme where, 
as we showed in Section~\ref{subsec:SMOMcons}, the value 1
is obtained. 
This result is shown by the black line at 1 in Fig.~\ref{fig:MOM-con}. 

To understand the discrepancy from 1 for $Z_V^{\text{cons}}$ 
in the RI$'$-MOM case, we fit the points shown in 
Fig.~\ref{fig:MOM-con} (including the correlations between them) to a form that allows for both discretisation effects 
and condensate contributions:  
\begin{eqnarray} \label{eq:MOM-con-fit}
  Z_V^{\mathrm{cons}}(\mathrm{MOM})(a,\mu) &=& 1 + \sum_{i=1}^5 c_{a^2\mu^2}^{(i)} (a\mu/\pi)^{2i} \\
  &&\hspace{-7.0em} + \sum_{i=1}^5 c_{\alpha a^2\mu^2}^{(i)} (a\mu/\pi)^{2i} \alpha_{\overline{\mathrm{MS}}}(1/a) + c_{\alpha}(\alpha_{\overline{\mathrm{MS}}}(\mu)/\pi)^4 \nonumber \\
  &&\hspace{-8.0em}+ \sum_{j=1}^5 c_{\mathrm{cond}}^{(j)} \alpha_{\overline{\mathrm{MS}}}(\mu)\frac{(1\ \mathrm{GeV})^{2j}}{\mu^{2j}} 
  \times [1 + c_{\mathrm{cond},a^2}^{(j)}(a\Lambda/\pi)^2] \,.\nonumber
\end{eqnarray}
Note that this constrains $Z_V^{\mathrm{cons}}(\mathrm{MOM})$ to be 1 
in the continuum once condensates are removed. Here
$\alpha_{\overline{\mathrm{MS}}}(\mu)$ is the value of the strong coupling constant 
in the $\overline{\mathrm{MS}}$ scheme at the scale
$\mu$ calculated from running the value obtained in \cite{Chakraborty:2014aca} using 
the four-loop QCD $\beta$ function. The
fit allows for discretisation errors of the generic form $(a\mu)^{2i}$ and terms 
$\mathcal{O}(\alpha_s(a\mu)^{2i})$; only even
powers of $a$ appear due to the remnant chiral symmetry of staggered quarks. 
Note that in principle we have removed $(a\mu)^2$ terms by dividing by 
$V^{\text{cons}}_{\gamma \otimes I}$; the fit returns only a small coefficient for this term.  
The $\alpha_s$-suppressed discretisation terms are included as the very
small statistical uncertainties on the results mean that these terms can 
have an effect in the fit. 
The
fourth term allows for systematic uncertainty from the missing $\alpha_s^4$ term 
in the RI$'$-MOM to $\overline{\mathrm{MS}}$ conversion factor (Eq.~\ref{eq:msbarmom}).

The condensate terms on the final line of Eq.~(\ref{eq:MOM-con-fit}) 
start at $1/\mu^2$ to allow for the gauge-noninvariant
$\langle A^2 \rangle$ condensate present in the operator product expansion (OPE) 
of the quark propagator~\cite{Chetyrkin:2009kh}. For the MOM kinematic
setup it is not possible to perform an OPE for the vertex functions as 
they are not short-distance quantities ($q=0$), so a complete analysis of what 
nonperturbative artefacts we expect to see in $Z_V$ is not possible. 
However, on general grounds we expect terms with inverse powers of $\mu$ to 
appear and we allow these terms also to have discretisation effects. 
We include even inverse powers of $\mu$ up to $1/\mu^8$. 

We use a Bayesian fit approach~\cite{Lepage:2001ym} in 
which coefficients are constrained by priors with 
a Gaussian distribution of a given central value and width. 
All coefficients in the fit form of Eq.~(\ref{eq:MOM-con-fit}) 
are given priors of $0 \pm 1$, except for that of the $(\alpha_s/\pi)^4$ term which 
has prior $0 \pm 5$ based on the lower-order coefficients. 
The choices for the priors are based on reasonable 
values for the coefficients of the terms in the fit. For example, 
discretisation effects are expected to appear as even powers 
of a physical scale (such as $\mu$ or $\Lambda$ here) divided by 
the ultraviolet cutoff ($\pi/a$) with coefficients of order one. 

The results of the fit are shown as the coloured dashed lines
in Fig.~\ref{fig:MOM-con}. 
The fit has a $\chi^2/\text{dof}$ of 0.6. It is already obvious from 
the figure that discretisation effects are not the only 
source of the discrepancy 
in $Z_V$ from 1. This is emphasised by attempting the fit without 
condensate terms (i.e. missing the last line of Eq.~(\ref{eq:MOM-con-fit})). 
Without the condensate terms
the quality of the fit is very poor, with a $\chi^2/{\text{dof}}$ of 7.7,  
in contrast to the fit of Eq.~(\ref{eq:MOM-con-fit}).
The sizeable contribution from the lowest order condensate is 
reflected in the coefficient found by the fit of
\begin{align} \label{eq:conserved-conds}
  &c_{\mathrm{cond}}^{(1)} = 0.154(54) .
\end{align}
The higher-order condensates cannot be pinned down by the fit. 

The correct answer for $Z_V$ for the conserved current in the 
continuum limit is, of course, 1. 
Our results and fit show that this can only be obtained from 
a calculation in the RI$'$-MOM scheme by working at multiple 
$\mu$ values at multiple values of the lattice spacing and fitting 
as a function of $\mu$ and $a$ to identify and remove 
the condensate contributions. If this is not done, systematic errors 
of $\mathcal{O}(1\%)$ (depending on the $\mu$ value) are  
present in $Z_V$, as is clear from Fig.~\ref{fig:MOM-con}. 

The issue will resurface when we discuss the use of the RI$'$-MOM scheme 
to renormalise nonconserved currents, specifically the HISQ local vector current, 
in Section~\ref{subsec:MOMloc}.

\subsection{$Z_V$ for the local current in the RI-SMOM scheme}
\label{subsec:SMOMloc}

\begin{figure}
  \includegraphics[width=0.47\textwidth]{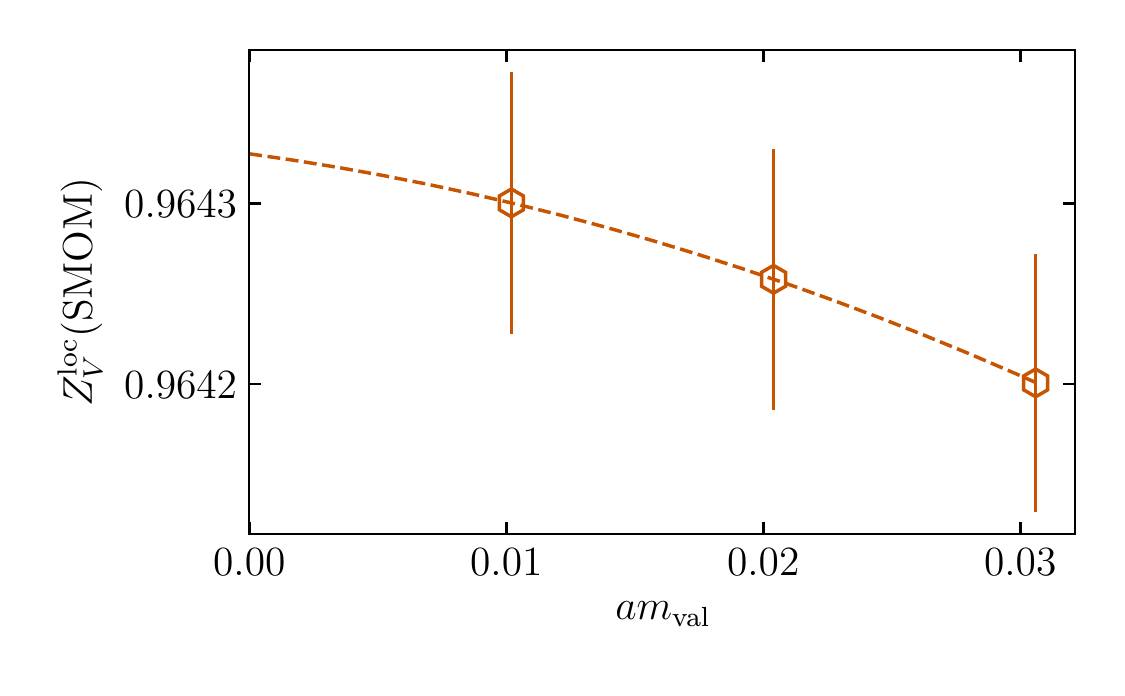}
  \caption{Valence mass dependence of $Z_V^{\text{loc}}(\text{SMOM})$ 
values obtained in the RI-SMOM scheme. 
  Results and extrapolation are shown for $\mu=3$ GeV on set 2. 
}
  \label{fig:mass-dep-loc-SMOM}
\end{figure}

We now turn to the calculation of the renormalisation constant 
for a nonconserved vector current using the RI-SMOM scheme. 
We will study the local current constructed from HISQ quarks since 
this is the simplest current and used in many analyses, such as the connected 
hadronic vacuum polarisation contribution to the anomalous magnetic moment 
of the muon~\cite{Chakraborty:2014mwa}.  

In~\cite{Chakraborty:2017hry} the renormalisation constant for the 
HISQ local current was calculated using the form factor method discussed 
in Section~\ref{sec:ward}. Results are given for very coarse, coarse and fine lattices 
in Table IV of that reference. The calculation was done using valence 
$s$ quarks and the form factor was determined for the local temporal vector current 
between two $s\overline{s}$ pseudoscalar 
mesons at rest\footnote{Note that the `spectator' quark used the clover formalism in this case, 
in order for the staggered tastes to cancel in the correlation function.}. 
From the discussion in Section~\ref{sec:ward} we expect such a determination 
of $Z_V$ to be exact so that $Z_V^{\text{loc}}(\text{F(0)})$ is equal 
to a perturbative series in $\alpha_s$ that matches the lattice scheme to the 
$\overline{\text{MS}}$ scheme, up to discretisation effects. This was tested 
in~\cite{Chakraborty:2017hry} (Appendix B) by fitting the $Z_V$ results to this 
form, including the known $\mathcal{O}(\alpha_s)$ coefficient in the perturbative 
series. A good fit was obtained that allowed values for $Z_V^{\text{loc}}(\text{F(0)})$ 
to be inferred on finer lattices.   
Here we will calculate $Z_V^{\text{loc}}(\text{SMOM})$ and compare it to 
$Z_V^{\text{loc}}(\text{F(0)})$. They should both contain the same 
perturbative series (since this is unique for a given operator)
and differ only by discretisation effects. 

\begin{table*}
\caption{Local vector current renormalisation factors, $Z_V^{\text{loc}}$ 
for a variety of $\mu$ values (given in column 2) on gluon field configurations at different 
lattice spacing values (denoted by the set number in column 1). 
Column 3 gives results using the RI-SMOM scheme and column 4 gives results 
using the standard RI$'$-MOM scheme. Note that the RI$'$-MOM results include the 
additional renormalisation factor of Eq.~(\ref{eq:msbarmom}) 
(Table~\ref{tab:RI-MSbar-conv}) that is needed to take the lattice 
current all the way to the $\overline{\text{MS}}$ scheme. Results are extrapolated to zero valence 
quark mass. Columns 5 and 6 give results for the modified (denoted by Rc) RI$'$-MOM and 
RI-SMOM$_{\gamma_{\mu}}$ schemes in which 
a ratio to the value for the conserved current
renormalisation in that scheme has been taken (Eq.~(\ref{eq:Rcdef})). 
}
\label{tab:local}
\begin{ruledtabular}
\begin{tabular}{llllll}
Set & $\mu$ [GeV] & $Z_V^{\mathrm{loc}}(\mathrm{SMOM})$ & $Z_V^{\mathrm{loc}}(\mathrm{MOM})$ & $Z_V^{\mathrm{loc}}(\mathrm{MOM}_{\text{Rc}})$ & $Z_V^{\mathrm{loc}}(\mathrm{SMOM}_{\gamma_{\mu},\text{Rc}}) $\\
\hline
1 & 1 & 0.9743(11) & - & - & - \\
2 & 1 & 0.9837(20) & - & - & - \\
\hline
1 & 2 & 0.95932(18) & - & - & - \\
2 & 2 & 0.97255(22) & 0.98771(85) & 0.97012(25) & 0.91864(25) \\
6 & 2 & 0.98445(11) & 0.99784(79) & 0.98292(44) & 0.959434(58) \\
7 & 2 & 0.99090(36) & 1.00202(89) & 0.99012(19) & 0.982435(21) \\
\hline
2 & 2.5 & 0.96768(12) & 0.97968(34) & 0.96447(17) & 0.89506(19) \\
\hline
2 & 3 & 0.964328(75) & 0.97434(26) & 0.96027(23) & 0.87733(21) \\
6 & 3 & 0.977214(35) & 0.98785(28) & 0.97608(14) & 0.930025(40) \\
7 & 3 & 0.98702(11) & 0.99651(43) & 0.98633(11) &  0.969563(42) \\
\hline
6 & 4 & 0.972415(18) & 0.98090(16) & 0.971009(90) & 0.905823(40) \\
7 & 4 & 0.983270(54) & 0.99241(21) & 0.982942(40) & 0.954992(30) \\
\end{tabular}
\end{ruledtabular}
\end{table*}

\begin{figure}
  \includegraphics[width=0.47\textwidth]{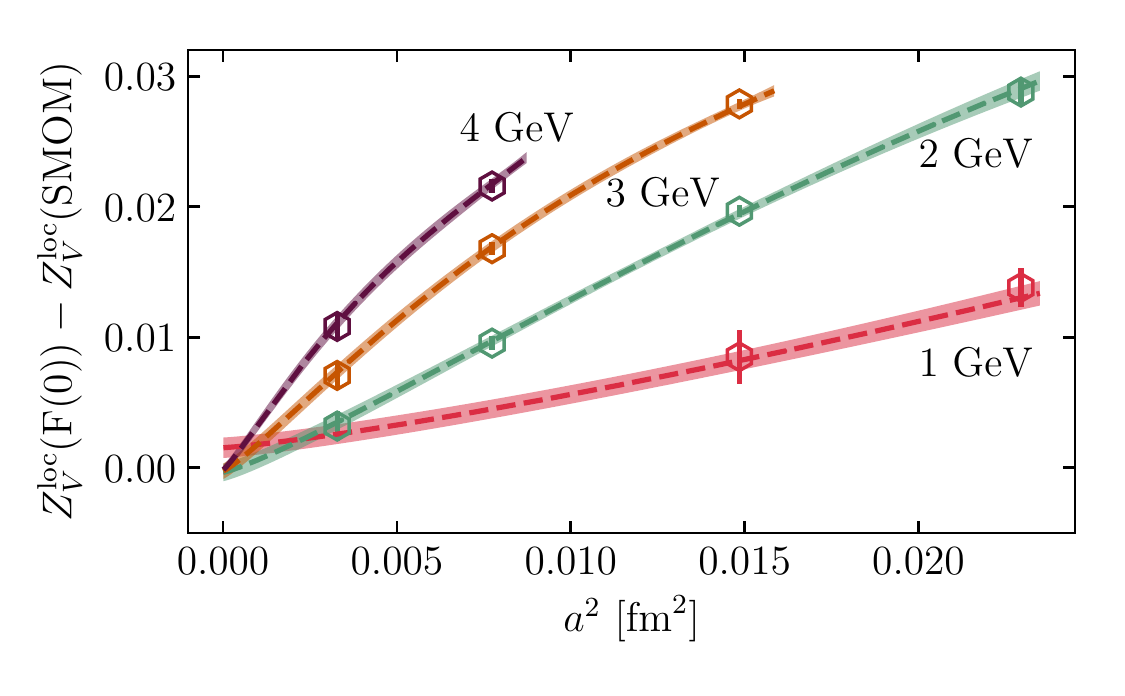}
  \includegraphics[width=0.47\textwidth]{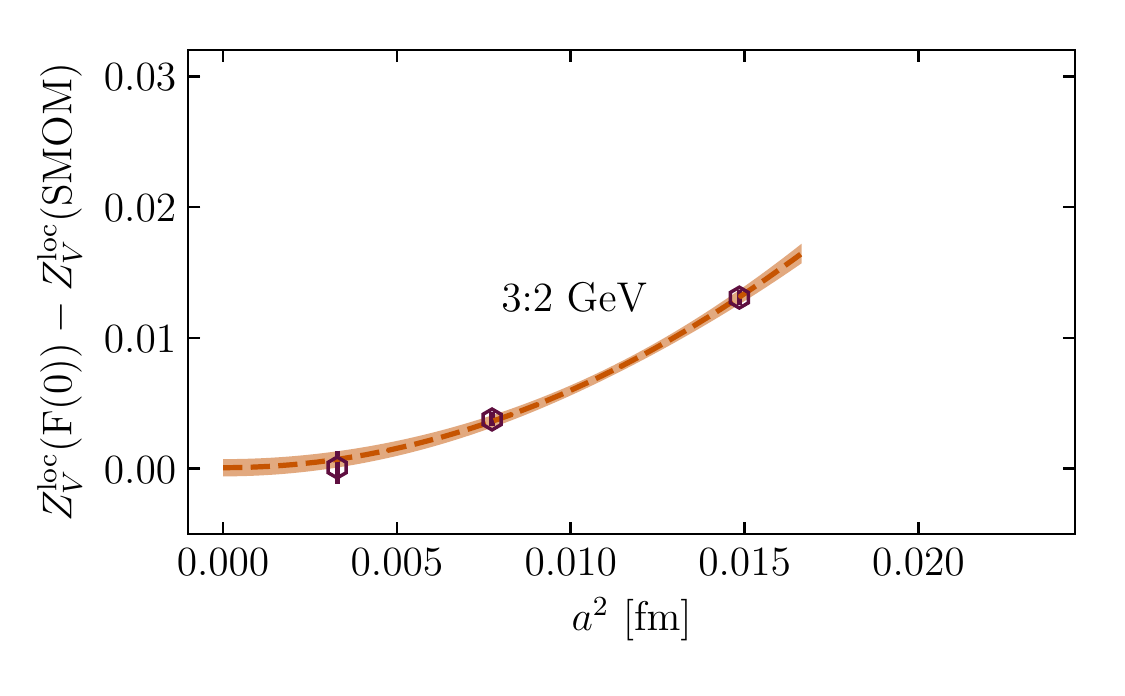}
  \caption{The top plot shows $Z_V^{\mathrm{loc}}(\mathrm{SMOM})$ for $\mu$ values between 1 GeV and 4 GeV, 
plotted as a difference to the corresponding $Z_V$ at that lattice spacing 
obtained from the vector form factor at 
zero momentum-transfer. The fit shown (see text) accounts for
  discretisation errors and condensate contributions, but no condensate contributions are seen 
and they are strongly constrained to zero by the fit. 
The lower plot shows the same difference but for a $Z_V^{\text{loc}}(\mathrm{SMOM})$ 
derived from results at $\mu$ = 2 GeV and 3 GeV in such a way as to reduce 
discretisation effects (see Eq.~(\ref{eq:zv-adjust})). 
The fit here is to a simple (constant + $a^4$) form as described in the text. 
  }
  \label{fig:SMOM-loc}
\end{figure}

To calculate $Z_V^{\text{loc}}(\text{SMOM})$ a little care is required 
in the construction of the SMOM vector vertex function with 
HISQ quarks. The operator
$\slashed{q}q_{\mu}\Lambda_V^{\mu}$ of Eq.~(\ref{eq:zvdef}) 
 must be constructed to be a taste singlet. For a local 
(in spin-taste notation, $\gamma_{\mu}\otimes \gamma_{\mu}$) current $\Lambda_V^{\mu}$ will 
have taste $\gamma_{\mu}$. This means that 
the $\slashed{q}$ in the vertex function must also have this taste. 
The correct construction
is as
\begin{equation} \label{eq:loc-SMOM}
  \sum_{\mu,\nu} \hat{q}_{\nu} \overline{\overline{(\gamma_{\nu} \otimes \gamma_{\mu})}} \hat{q}_{\mu} \Lambda_{V,\text{loc}}^{\mu} .
\end{equation}
Taking the spin-colour trace of this operator and dividing by 
$48\hat{q}^2$ then gives $Z_q/Z_V$. The wavefunction renormalisation 
is calculated in the same way as for the conserved current, Eq.~(\ref{eq:zqdefhisq}). 
Results for $Z_V^{\text{loc}}(\text{SMOM})$ are given in Table~\ref{tab:local} (column 3). 
This is after extrapolation to zero valence quark mass. Figure~\ref{fig:mass-dep-loc-SMOM}
shows that the impact of this is very small 
(we expect in this case that the mass dependence is purely 
a discretisation effect).

Figure~\ref{fig:SMOM-loc} (top plot) shows our results as a difference between 
$Z_V^{\text{loc}}({\text{SMOM}})$ and $Z_V^{\text{loc}}(\text{F(0)})$. 
$Z_V^{\text{loc}}(\text{F(0)})$ values are from~\cite{Chakraborty:2017hry} 
and obtained on the same gluon field configurations that we use here. 
We plot the difference for the multiple $\mu$ values used for the 
$Z_V^{\text{loc}}(\text{SMOM})$ determination as a function of lattice spacing 
in Fig.~\ref{fig:SMOM-loc}. Results are shown from very coarse to superfine 
lattice spacings noting that higher $\mu$ values are only accessible on 
finer lattices because of the constraint in Eq.~(\ref{eq:redbrill}).   

We can readily fit this difference of $Z_V^{\text{loc}}$ values, 
$\Delta Z_V^{\text{loc}}$, to a function constructed from possible
discretisation effects. To keep the fit as general as possible we 
also allow for the existence of condensate terms to see to what 
extent they are constrained 
by the fit.  We also allow for condensate terms multiplied by discretisation 
effects that would vanish in the continuum limit (and are therefore benign). 
We use
\begin{eqnarray} \label{eq:fitzvdiff-disc}
  \Delta Z_V^{\mathrm{loc}}(a,\mu) &=& \sum_{i=1}^{3} \left[ c_{a^2\mu^2}^{(i)} (a\mu/\pi)^{2i} \right.\\
 && \left. \hspace{2.0em}+ c_{\alpha a^2\mu^2}^{(i)} (a\mu/\pi)^{2i} \alpha_{\overline{\mathrm{MS}}}(1/a) \right] \nonumber \\
  &&\hspace{-8.0em}+ \sum_{j=1}^{3} c_{\mathrm{cond}}^{(j)} \alpha_{\overline{\mathrm{MS}}}(\mu)\frac{(1\ \mathrm{GeV})^{2j}}{\mu^{2j}} 
  \times [1 + c_{\mathrm{cond},a^2}^{(j)}(a\Lambda/\pi)^2] \,.\nonumber
\end{eqnarray}
All coefficients are given priors $0\pm 1$. 
This fit has a $\chi^2/\text{dof}$ value of 0.18 and 
finds no significant condensate contribution. The lowest order ($1/\mu^2$) condensate term 
is constrained by the fit to have a very small coefficient compatible with zero: -0.020(44) 
(compare. Eq.~(\ref{eq:conserved-conds})). Thus we see that $\Delta Z_V^{\text{loc}}$ 
is compatible with being, as expected, purely a discretisation effect.  

We have shown here that the $Z_V$ value obtained for the nonconserved local 
HISQ current using the RI-SMOM scheme is indeed exact i.e. it has no 
nonperturbative condensate contributions (visible at our high level of accuracy) 
that would survive the continuum limit 
as a source of systematic error. This can be traced to the fact that the 
condensate contributions present in the vector vertex function for 
the conserved vector current and in the 
inverse propagator must cancel because of the Ward-Takahashi identity. 
This identity also protects $Z_V$ from any effects 
arising from the gauge-fixing procedure.

This means that there is in fact no lower limit in principle to the $\mu$ value 
that can be used for the vector current renormalisation in the RI-SMOM scheme. 
In Fig.~\ref{fig:SMOM-loc} (top plot) we include values corresponding to $\mu$ = 1 GeV. 
These show smaller discretisation effects than those for the higher $\mu$ values 
and so may be preferable on these grounds if only one $\mu$ value is used 
(which is all that is necessary in principle since no allowance needs to be 
made for condensate effects). 
The statistical errors possible with 20 configurations grow as $\mu$ is 
reduced. However, for $\mu = 1$~GeV the uncertainties could still readily be reduced to the 
0.1\% level with higher statistics. 

Smaller discretisation effects are possible by extrapolating in 
$\mu$ to $\mu=0$. 
A simple method that removes $\mu^2a^2$ terms in 
$Z_V^{\text{loc}}(\text{SMOM})$ combines results at two 
different $\mu$ values (for a given lattice spacing) to determine a new value 
\begin{eqnarray}
\label{eq:zv-adjust}
Z_V^{\text{loc}}(\text{SMOM})(\mu_1,\mu_2) &=& \\
&&\hspace{-5.0em} \frac{\mu^2_1 Z_V^{\text{loc}}(\text{SMOM})(\mu_2) - \mu_2^2 Z_V^{\text{loc}}(\text{SMOM})(\mu_1)}{\mu^2_1-\mu^2_2} \, . \nonumber
\end{eqnarray}
This can always be done, given that $Z_V^{\text{loc}}(\text{SMOM})$ only depends 
on $\mu$ through discretisation effects. 
We use $\mu_1$ = 3 GeV and $\mu_2$ = 2 GeV and Eq.~(\ref{eq:zv-adjust}) returns 
a precise result because the statistical uncertainties 
are very small on these $\mu$ values. 
We show the results of taking a difference 
to $Z_V^{\text{loc}}(\text{F(0)})$ for this new $Z_V$ value 
in the lower plot of Fig.~\ref{fig:SMOM-loc}. 
The points clearly have 
smaller discretisation effects compared to the original $Z_V$ values that they were derived 
from. Given that the discretisation effects in 
$Z_V^{\text{loc}}(\text{F(0)})$ were relatively small~\cite{Chakraborty:2017hry} 
we interpret this as a reduction 
of discretisation effects in $Z_V^{\text{loc}}(\text{SMOM})$. We can fit the points 
in the lower plot of Fig.~\ref{fig:SMOM-loc} to a very 
simple curve : $C + D(a \times 1 {\text{GeV}})^4$ and $C$ is found to be 0.00008(66).   
The smaller discretisation effects seen using Eq.~(\ref{eq:zv-adjust}) may make 
this approach preferable to that of using $Z_V^{\text{loc}}(\text{SMOM})$ for a 
single $\mu$ value although it doubles the cost. Using three values 
of $\mu$ a higher-order scheme could obviously be devised 
to reduce discretisation effects further. 

\subsection{$Z_V$ for the local current in the RI$'$-MOM scheme}
\label{subsec:MOMloc}

\begin{figure}
  \includegraphics[width=0.47\textwidth]{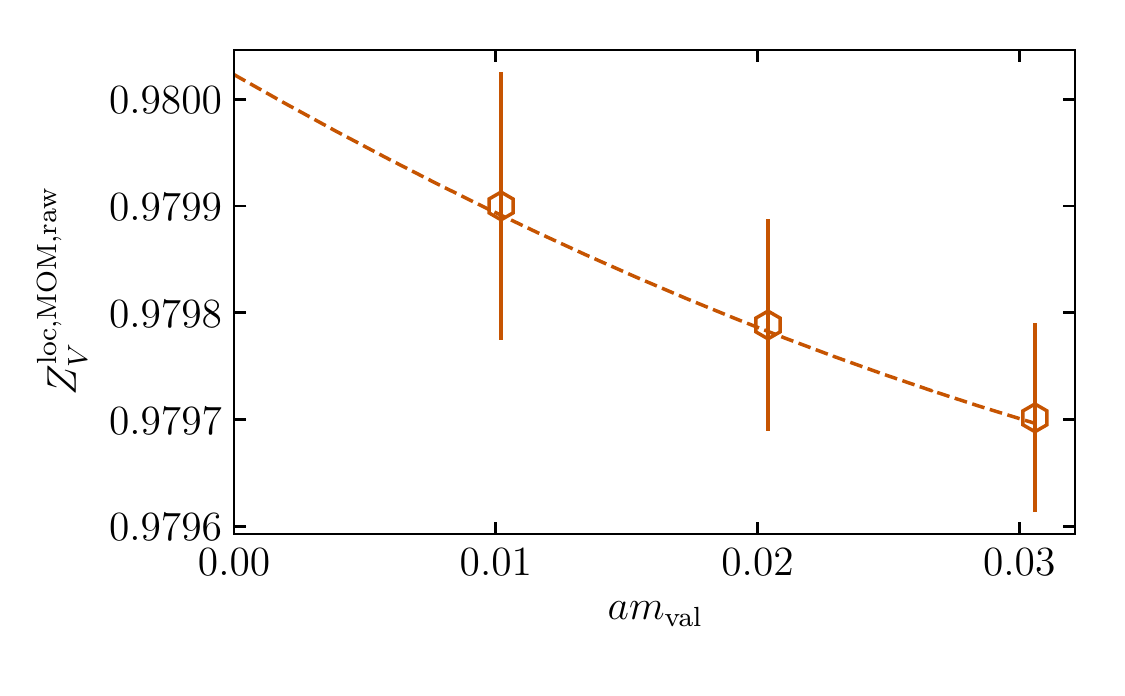}
  \caption{Valence mass dependence of our raw results for 
$Z_V^{\text{loc}}$ calculated in the RI$'$-MOM scheme, before 
multiplication by the additional renormalisation factor needed 
to match to $\overline{\text{MS}}$. 
  Results and extrapolation are shown for $\mu=3$ GeV on set 2. 
}
  \label{fig:mass-dep-loc-MOM}
\end{figure}

\begin{figure}
  \includegraphics[width=0.47\textwidth]{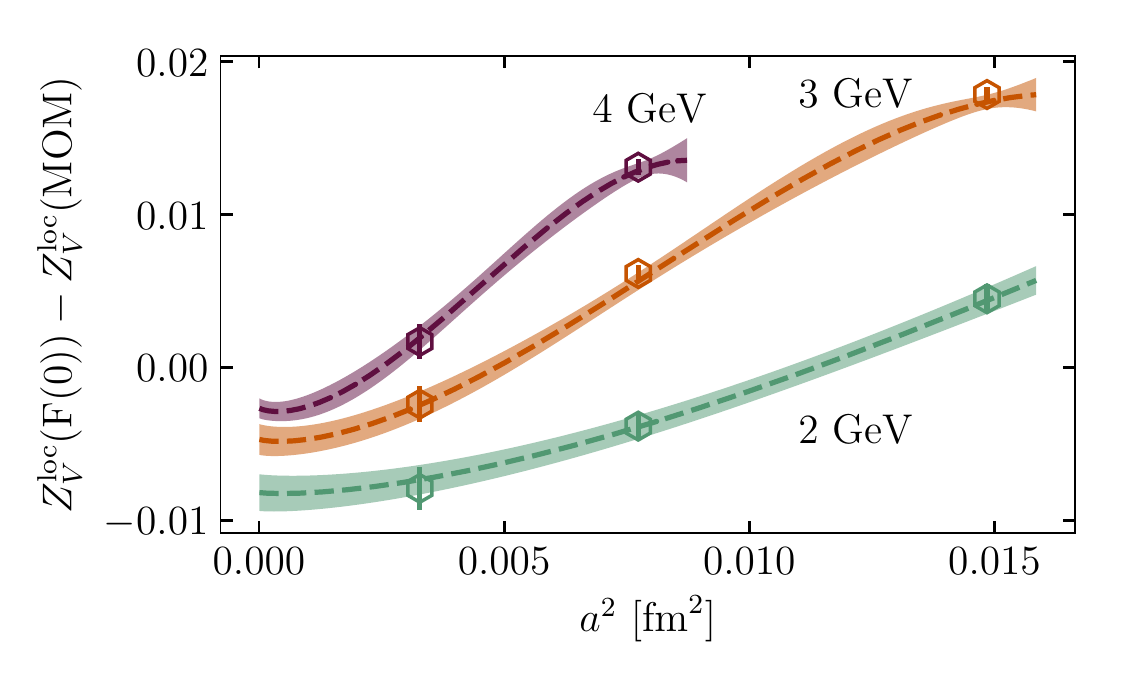}
  \caption{$Z_V^{\mathrm{loc}}(\mathrm{MOM})$ for $\mu$ values between 2 GeV and 4 GeV, 
plotted as a difference to the corresponding $Z_V$ at that lattice spacing 
obtained from the vector form factor at 
zero momentum-transfer. The fit shown (see text) accounts for
  discretisation errors and condensate contributions, with condensate contributions being 
necessary to obtain a good fit. 
  }
  \label{fig:MOM-loc}
\end{figure}

We now turn to the determination of the renormalisation constant for 
the nonconserved local vector current using the RI$'$-MOM scheme, 
$Z_V^{\text{loc}}(\text{MOM})$. Again, the vector vertex function must be 
a taste-singlet. The RI$'$-MOM scheme uses a simple $\gamma_{\mu}$ 
projector (Eq.~(\ref{eq:zvmomdef})), which for the HISQ local vector current needs to have 
spin-taste $\gamma_{\mu}\otimes \gamma_{\mu}$. Then we use 
\begin{equation} \label{eq:loc-MOM}
  \sum_{\mu} \overline{\overline{(\gamma_{\mu} \otimes \gamma_{\mu})}} \Lambda_{V,\text{loc}}^{\mu} .
\end{equation}
to determine $Z_V/Z_q$ along with Eq.~(\ref{eq:zqdefhisq}) to determine $Z_q$. 
Figure~\ref{fig:mass-dep-loc-MOM} shows the valence mass extrapolation for 
one set of raw results. Despite having a more significant mass extrapolation 
than for the RI-SMOM results (Fig.~\ref{fig:mass-dep-loc-SMOM}), 
this is still very mild. 
Table~\ref{tab:local} gives our results in column 4, where we note that the values given 
for $Z_V^{\text{loc}}(\text{MOM})$ include the additional renormalisation 
factor shown in Eq.~(\ref{eq:msbarmom}) and given in Table~\ref{tab:RI-MSbar-conv}. 

Figure~\ref{fig:MOM-loc} shows our results given, following the discussion 
in Section~\ref{subsec:SMOMloc}, as a difference to the renormalisation constants 
obtained for the local current using the form factor method in~\cite{Chakraborty:2017hry}. 
This figure is very different from Fig.~\ref{fig:SMOM-loc}, with the results showing no 
sign of converging to zero in the continuum limit that would demonstrate agreement between 
the form factor and RI$'$-MOM schemes for $Z_V$. This
shows the presence of condensate contributions in $Z_V^{\text{loc}}(\text{MOM})$ and 
to fit these results we need to include condensates that survive the continuum limit 
in the fit form. 

For the difference of $Z_V^{\text{loc}}$ values shown in Fig.~\ref{fig:MOM-loc} 
we use  the same fit form as that used earlier for the RI-SMOM results in 
Eq.~(\ref{eq:fitzvdiff-disc}) (with the addition of an $\alpha_s^4$ to allow for 
uncertainty in the matching from MOM to $\overline{\text{MS}}$ as used in 
Eq.~(\ref{eq:MOM-con-fit}); this term has very little effect). 
This fit, with $\chi^2/\text{dof} = $0.14 is 
shown by the dashed lines in Fig.~\ref{fig:MOM-loc}. 
It returns a coefficient for the leading-order condensate term of -0.209(63) which is 
consistent with the leading-order condensate term seen in 
the conserved current $Z_V$ calculated in the RI$'$-MOM scheme (Eq.~(\ref{eq:conserved-conds}), 
with opposite sign because of our definition of $\Delta Z_V$ here).  
Note the difference with the results in the RI-SMOM case. 

The results of Fig.~\ref{fig:MOM-loc} show that 
the standard RI$'$-MOM scheme cannot be used to determine an accurate result for 
$Z_V$ for nonconserved currents. If no attention is paid to the contamination of 
$Z_V$ by condensate contributions then $\mathcal{O}(1\%)$ systematic errors 
will be made.  

\begin{figure}
  \includegraphics[width=0.47\textwidth]{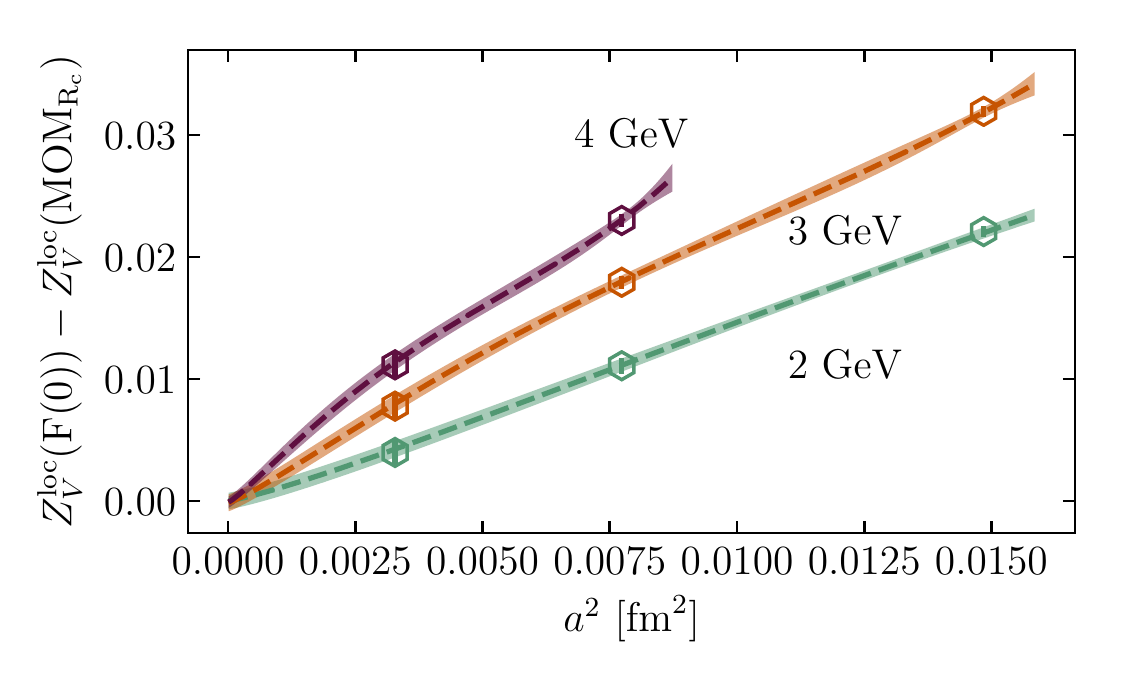}
  \caption{$Z_V^{\mathrm{loc}}(\mathrm{MOM}_{\text{Rc}})$ from the modified RI$'$-MOM scheme 
of Eq.~\ref{eq:Rcdef}, 
plotted as a difference to the corresponding $Z_V$ at that lattice spacing 
obtained from the vector form factor at 
zero momentum-transfer. Results are shown for $\mu$ values from 2 GeV to 4 GeV. 
The fit shown (see text) accounts for
  discretisation errors and condensate contributions, but condensate contributions 
are strongly constrained to be zero. 
  }
  \label{fig:MOMrat-loc}
\end{figure}
 
We can modify the RI$'$-MOM scheme to address this issue, however. 
We know that the conserved current and the renormalised local current are the 
same operator in the continuum limit and so their vertex functions 
must contain the same nonperturbative 
contributions from the RI$'$-MOM scheme in that limit. We can therefore calculate 
$Z_V^{\text{loc}}(\text{MOM})$ by taking a ratio of the vertex functions of 
the local and conserved currents. We call this scheme the RI$'$-$\text{MOM}_{\text{Rc}}$ 
scheme. Specifically we calculate 
\begin{equation}
\label{eq:Rcdef}
Z_V^{\text{loc}}(\text{MOM}_{\text{Rc}}) = \frac{\Tr(\gamma_{\mu}\Lambda^{\mu}_{V,\text{cons}})}{\Tr(\gamma_{\mu}\Lambda^{\mu}_{V,\text{loc}})} = \frac{Z_V^{\text{loc}}(\text{MOM})}{Z_V^{\text{cons}}(\text{MOM})} .
\end{equation}
Taking the ratio also means that 
no additional renormalisation is needed in this case. 

Our results from implementing this scheme are given in Table~\ref{tab:local} (column 5).
Figure~\ref{fig:MOMrat-loc} shows the results given once again as a difference to 
the renormalisation constant obtained for the local current in the form factor 
method.  We now see that the difference of $Z_V$ values clearly approaches 0 in the 
continuum limit and there is no sign of condensate contamination in that limit. 
The results in the RI$'$-MOM$_{\text{Rc}}$ scheme 
look very similar to those in the RI-SMOM scheme (see Fig.~\ref{fig:SMOM-loc}). 
We can fit the values for $\Delta Z_V^{\text{loc}}$ in Fig.~\ref{fig:MOMrat-loc} to 
the same form as that used for the RI-SMOM results (Eq.~(\ref{eq:fitzvdiff-disc})). 
The fit gives $\chi^2/\text{dof}$ = 0.32 and constrains the lowest-order condensate 
coefficient that would survive the continuum limit to -0.01(5). 

We conclude that the modified RI$'$-MOM scheme of Eq.~(\ref{eq:Rcdef})
does provide a method to determine an accurate renormalisation for the local vector current. 
The method does require calculations with the conserved current and so is more 
complicated than the RI-SMOM scheme. 

\subsection{$Z_V$ for the local current in the RI-SMOM$_{\gamma_{\mu}}$ scheme}
\label{subsec:SMOMgloc}

\begin{figure}
  \includegraphics[width=0.47\textwidth]{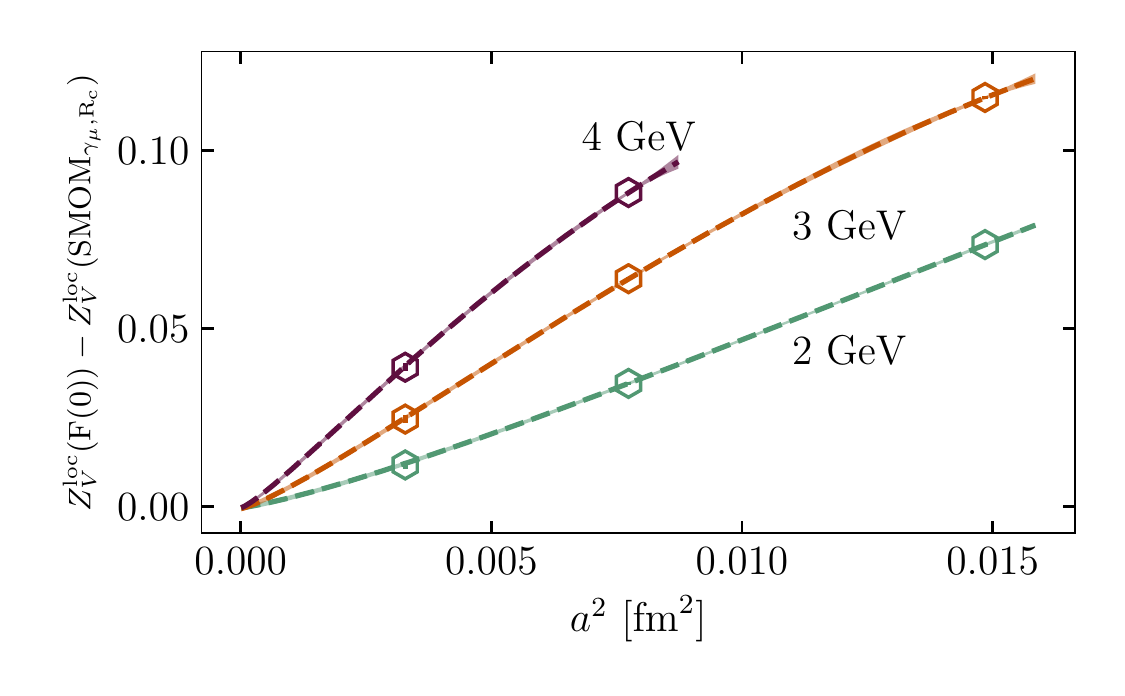}
  \caption{$Z_V^{\mathrm{loc}}(\mathrm{SMOM}_{\gamma_{\mu},\text{Rc}})$ from 
the modified RI-SMOM$_{\gamma_{\mu}}$ scheme, 
plotted as a difference to the corresponding $Z_V$ at that lattice spacing 
obtained from the vector form factor at 
zero momentum-transfer. Results are shown for $\mu$ values from 2 GeV to 4 GeV. 
The fit shown (see text) accounts for
  discretisation errors and condensate contributions, but condensate contributions 
are strongly constrained to be zero. 
  }
  \label{fig:SMOMgam-loc}
\end{figure}

An alternative momentum-subtraction scheme is the RI-SMOM$_{\gamma_{\mu}}$ scheme
which uses the same vertex function (and wavefunction renormalisation) as 
the RI$'$-MOM scheme but uses RI-SMOM kinematics 
(i.e. $q=p_1-p_2 \ne 0$, $p_1^2=p_2^2=q^2=\mu^2$). 

To obtain an accurate result for $Z_V$ for the local current (as an example 
of a nonconserved current) we must modify the scheme as was done for 
the RI$'$-MOM scheme in Eq.~(\ref{eq:Rcdef}). The only difference is 
that we must also modify the tree-level vertex function factor for 
the conserved current from that of Eq.~(\ref{eq:constree}) to 
reflect the SMOM kinematics. Table~\ref{tab:local} 
gives our results from this modified RI-SMOM$_{\gamma_\mu,\text{Rc}}$ scheme 
in column 6. Figure~\ref{fig:SMOMgam-loc} plots the difference of these 
$Z_V$ values with those from using the form factor method. 
We see that, as for the SMOM scheme in Fig.~\ref{fig:SMOM-loc} and the
modified RI$'$-MOM scheme in Fig.~\ref{fig:MOMrat-loc}, 
the values converge to zero as $a\rightarrow 0$ as discretisation 
effects should. Discretisation effects are significantly larger 
here than in the previous schemes, however. We fit the results to the same functional 
form as used for the other schemes (i.e.\ Eq.~(\ref{eq:fitzvdiff-disc})) 
and obtain a good fit (we double the prior width on $(a\mu)^n$ terms 
to allow for the larger 
discretisation effects). 
$\chi^2/\text{dof}$ = 0.56 and the lowest order 
condensate coefficient is constrained very tightly, as in the other 
exact cases, to -0.03(5). 

The same conclusions apply as for the RI$'$-MOM scheme, i.e. that 
defining $Z_V$ from the ratio of vertex functions with the conserved 
current gives an exact result. 

\subsection{Renormalisation of the axial vector current}
\label{subsec:ZA}

\begin{figure}
  \includegraphics[width=0.47\textwidth]{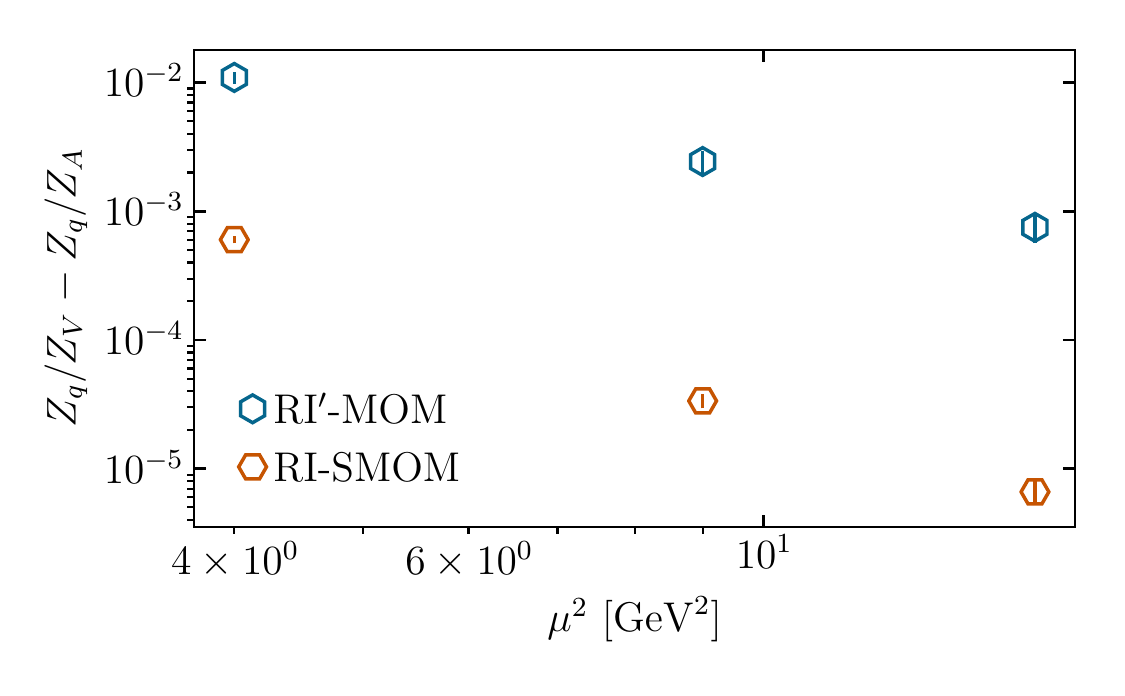}
  \caption{The difference between the vertex functions for the local vector and 
local axial vector currents as a function of $\mu$ in the RI$'$-MOM
  and RI-SMOM schemes. The RI-SMOM scheme gives a much smaller nonperturbative 
contribution to $Z_A-Z_V$ than the RI$'$-MOM scheme, but neither scheme gives 
$Z_V=Z_A$ which is true to all orders in perturbation theory.  
The values plotted result 
from an extrapolation to zero valence quark mass
  and are shown for the finest lattice we use (set 7 in Table \ref{tab:ensembles}).
  }
  \label{fig:VminusA}
\end{figure}

The renormalisation factors for axial vector currents can also be calculated 
using momentum-subtraction schemes. However, for actions with sufficient chiral 
symmetry the axial vector current renormalisation, $Z_A$, can be related to the 
vector current renormalisation at zero quark mass. 
For example, for staggered quarks, $Z_{S\otimes T}=Z_{S5\otimes T5}$ to all 
orders in perturbation theory~\cite{Sharpe:1993ur}. Here $S\otimes T$ indicates 
the operator spin-taste and $S5=\gamma_5S$. This means that the local axial 
vector current and local vector current have the same renormalisation factor. 

Having shown that the local vector current renormalisation factor can be 
calculated accurately and without contamination by condensate contributions
in the RI-SMOM scheme in Section~\ref{subsec:SMOMloc}, 
it therefore makes sense to use this value also 
for the local axial vector current. 
Indeed, doing a separate calculation of $Z_A^{\text{loc}}$
risks introducing condensate contributions where none would be found using 
$Z_A=Z_V$.  Figure~\ref{fig:VminusA} shows the difference between 
the local vector and local axial vector vertex functions after 
extrapolation to zero quark mass, on the superfine lattices, set 7. 
Each point plotted is the difference of the local vector and local 
axial vector vertex functions i.e. $Z_q/Z_V-Z_q/Z_A$. 

We see that the difference in the RI-SMOM scheme is small but not 
zero. The results demonstrate approximately $\mu^{-6}$ behaviour 
expected on the basis of a chiral symmetry breaking condensate 
contribution~\cite{Aoki:2007xm}. Note that this contribution comes 
from $Z_q/Z_A$. 
For the RI$'$-MOM scheme the difference is much larger then for RI-SMOM and has a 
smaller slope in this log-log plot. This reflects 
the known impact of chiral symmetry breaking nonperturbative artefacts 
in this scheme~\cite{Aoki:2007xm}.   
In both cases it would be preferable to use $Z_A=Z_V$, in the RI$'$-MOM 
case using the modified RI$'$-MOM$_\text{Rc}$ approach of Eq.~(\ref{eq:Rcdef}). 

\section{Including quenched QED effects}
\label{sec:QED}

As lattice QCD calculations reach sub-percent precision it will become necessary to evaluate the 
electromagnetic corrections expected at this level. If QED effects are included in 
calculations involving nonconserved vector 
currents, such as the ongoing Fermilab/HPQCD/MILC calculations of the hadronic vacuum polarisation contribution 
to the anomalous magnetic moment of the muon~\cite{Davies:2019efs}, then consistency requires 
that QED effects are also included in the vector current renormalisation. Here we will 
study the impact of the valence quarks having electric charge on the renormalisation of the local vector current 
using the RI-SMOM scheme (for earlier results using different methods 
see~\cite{Boyle:2017gzv, Giusti:2019xct}). 

We include `quenched QED' in our lattice calculations by multiplying our QCD gauge fields by a 
U(1) gauge field representing the photon. The photon field, $A_{\mu}(k)$, is randomly generated in momentum space 
from a Gaussian distribution with variance $1/\hat{k}^2$ to yield the correct $\mathcal{O}(a^2)$-improved 
Feynman gauge propagator on the lattice (the definition of $\hat{k}$ is 
given in Eq.~(\ref{eq:qhatdef})). $A_{\mu}(k)$ is then converted to Landau gauge 
and transformed to position space. To make sure of the correct gauge covariance in position 
space it is important to remember that the position of the gauge fields is at the centre 
of the links, and not the sites~\cite{Drummond:2002yg}. The $A_{\mu}$ field in position 
space is then used as the phase  to construct
a U(1) field~\cite{Duncan:1996xy} in the form $\exp(ieQA_{\mu})$ where 
$Q$ is the charge of the quark that will interact with the field, in 
units of the charge on the proton, $e$. 
We use the 
$\mathrm{QED}_L$ formulation of compact QED~\cite{Hayakawa:2008an}, in which all zero modes are set to zero, 
$A_{\mu}(k_0,\mathbf{k}=0)=0$ with $A_{\mu}$ in Landau gauge 
(for a review of approaches to handling zero modes 
in QED on the lattice see~\cite{Patella:2017fgk}). 
We multiply the gluon field for each link of the lattice 
by the appropriate U(1) field before applying the HISQ smearing. 
The valence quarks can then interact with 
the photon via the standard HISQ action. 
Note that the sea quarks remain electrically neutral, so this is not a fully realistic scenario. 
Nevertheless it allows us to evaluate the most important QED effects. 

\begin{figure}
  \includegraphics[width=0.47\textwidth]{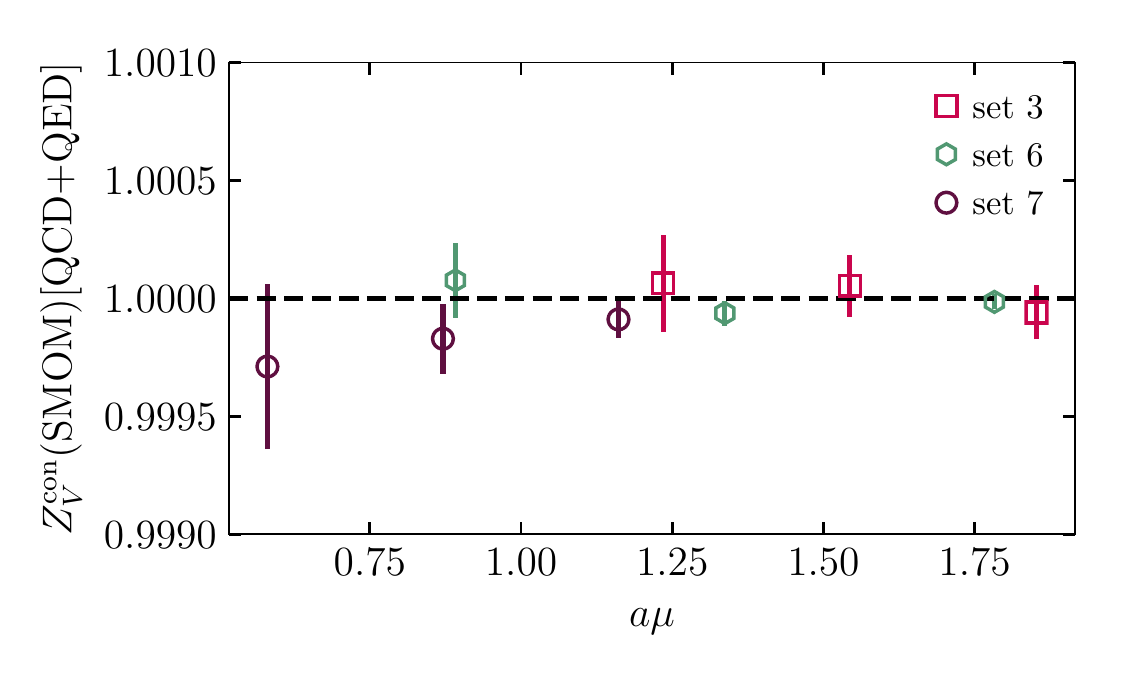}
  \caption{Results for the renormalisation factor, $Z_V$ for the QCD+QED conserved current 
for the HISQ action, calculated using the RI-SMOM scheme. 
Results are given for coarse, fine and superfine gluon field 
configurations for quark electric charge, $Q=2e/3$ and a variety of momenta with 
magnitude $a\mu$ in lattice units. 
}
  \label{fig:SMOM-con-u1}
\end{figure}

We have 
tested that the U(1) configurations we generate correctly reproduce 
the $\mathcal{O}(\alpha_{\mathrm{QED}})$ 
perturbation theory prediction for the average plaquette~\cite{Portelli:2010yn}, independent 
of gauge choice. Our results for the average value of the U(1) link field also 
agree with the $\mathcal{O}(\alpha_{\mathrm{QED}})$ expectations:  
\begin{eqnarray}
\label{eq:u1link}
\text{Landau \, gauge} \, &:& 1-\alpha_{\text{QED}}Q^2 0.0581   \\
\text{Feynman \, gauge} \, &:& 1-\alpha_{\text{QED}}Q^2 0.0775     \nonumber  
\end{eqnarray}
Note that the Landau gauge $\mathcal{O}(\alpha_{\text{QED}}Q^2)$ coefficient is 
$1/C_F=3/4$ that of the corresponding QCD result for the 
$a^2$-improved gluon action~\cite{Hart:2004jn} 
since the gluon propagator then has the same form as that of the photon here.  
The Feynman gauge coefficient is then 4/3 of the Landau gauge coefficient. 

Although we have tested calculations as a function of quark charge, $Q$, 
the results we will show here are all for 
$Q=2/3$. 
The results are not extrapolated to zero 
valence quark mass and are instead just the values at the sea light 
quark mass on each ensemble. The valence mass 
dependence of the results is observed to be negligibly small, as was the case in pure QCD.

An important test of the interaction between the quarks and 
the QCD+QED gauge fields  
is that $Z_V=1$ for the QCD+QED conserved current in the 
RI-SMOM scheme, as expected from a trivial extension of the 
Ward-Takahasi Identity to 
this case. This is demonstrated in Fig.~\ref{fig:SMOM-con-u1}.  

Our analysis for the renormalisation of the local vector current in the RI-SMOM scheme will 
study the ratio of the $Z_V^{\mathrm{loc}}$ calculated with and without 
the inclusion of electromagnetic effects. We proceed exactly as for the pure QCD case 
discussed in Section~\ref{subsec:SMOMloc}. The strong 
correlations between the QCD and QCD+QED calculations allow 
very precise determination of this ratio
(a typical correlation being $\sim 0.99$). 
We will denote a quantity $X$ 
calculated in pure QCD as $X[\mathrm{QCD}]$ while the same quantity 
calculated with the inclusion of QED effects 
will be denoted $X[\mathrm{QED+QCD}]$. We will also employ 
the notation $X[\mathrm{(QCD+QED)/QCD}] \equiv X[\mathrm{QED+QCD}]/X[\mathrm{QCD}]$.  

\begin{figure}
  \includegraphics[width=0.47\textwidth]{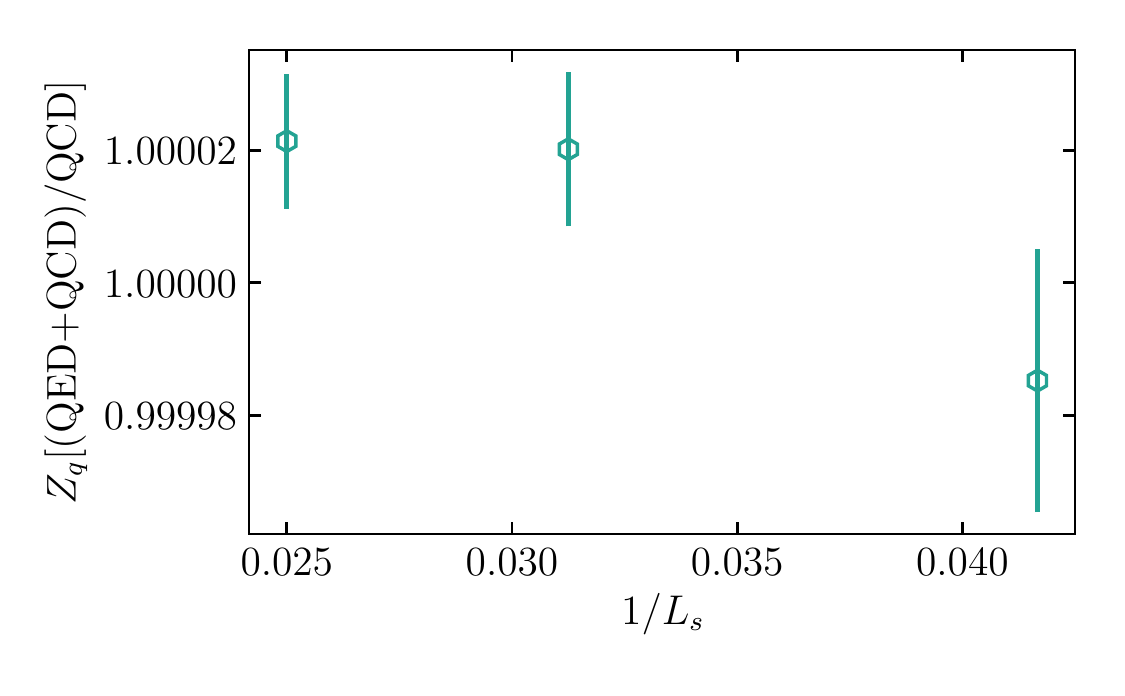}
  \includegraphics[width=0.47\textwidth]{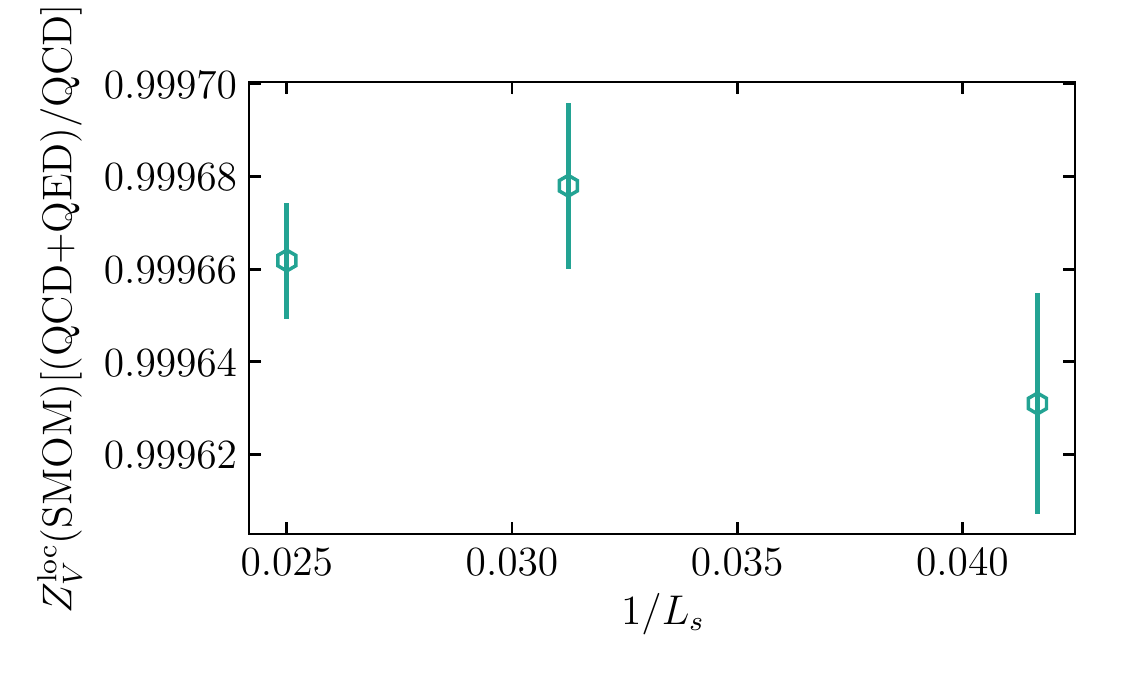}
  \caption{The impact of quenched QED (with quark charge $2e/3$) 
on the determination of $Z_V^{\text{loc}}$ and $Z_q$ using 
the RI-SMOM scheme as a function of the lattice spatial extent, $L_s$ 
in lattice units. Results are for coarse lattices, sets 3, 4 and 5, 
and $\mu$ = 2 GeV. The volume dependence is negligible.  
}
  \label{fig:volume}
\end{figure}

Because QED is a long-range interaction it is important to test 
finite-volume effects, although we do not expect them to be 
large here since we studying renormalisation of electrically neutral 
currents. 
The finite-volume effects in the self-energy function of fermions 
has been studied in \cite{Davoudi:2018qpl} with the result that for 
off-shell quarks the finite-volume effects start at order $1/L_s^2$ 
where $L_s$ is lattice spatial extent. 
This implies that even the finite-volume effects 
for quantities such as $Z_q$ should be small. 
Figure~\ref{fig:volume} confirms both of these expectations with results on 
the three lattice sets with the same parameters but different volumes 
(sets 3, 4 and 5, ranging in spatial extent from 2.9 fm to 4.9 fm). 
Negligible effects are seen here
and we therefore ignore finite-volume issues in the following analysis.  

\begin{table}
\caption{The ratio of renormalisation factors $Z_V$ 
for the QCD + quenched QED case 
to the pure QCD case. These are for the local HISQ vector current 
calculated in the RI-SMOM scheme on gluon field configuration sets 
listed in column 1 and at $\mu$ values listed in column 2 (and at a valence 
quark mass of $m_l$). 
}
\label{tab:ZV-qed-data}
\begin{ruledtabular}
\begin{tabular}{lll}
Set & $\mu$ [GeV] & $Z_V^{\mathrm{loc}}(\mathrm{SMOM})[\mathrm{(QED+QCD)/QCD}]$\\
\hline
3 & 2 & 0.999631(24) \\
6 & 2 & 0.999756(32) \\
7 & 2 & 0.999831(43) \\
\hline
3 & 2.5 & 0.999615(12) \\
\hline
3 & 3 & 0.999622(13)  \\
6 & 3 & 0.9997043(39)  \\
7 & 3 & 0.9997797(92)  \\
\hline
6 & 4 & 0.9996754(26) \\
7 & 4 & 0.9997425(24) \\
\end{tabular}
\end{ruledtabular}
\end{table}

\begin{figure}
  \includegraphics[width=0.47\textwidth]{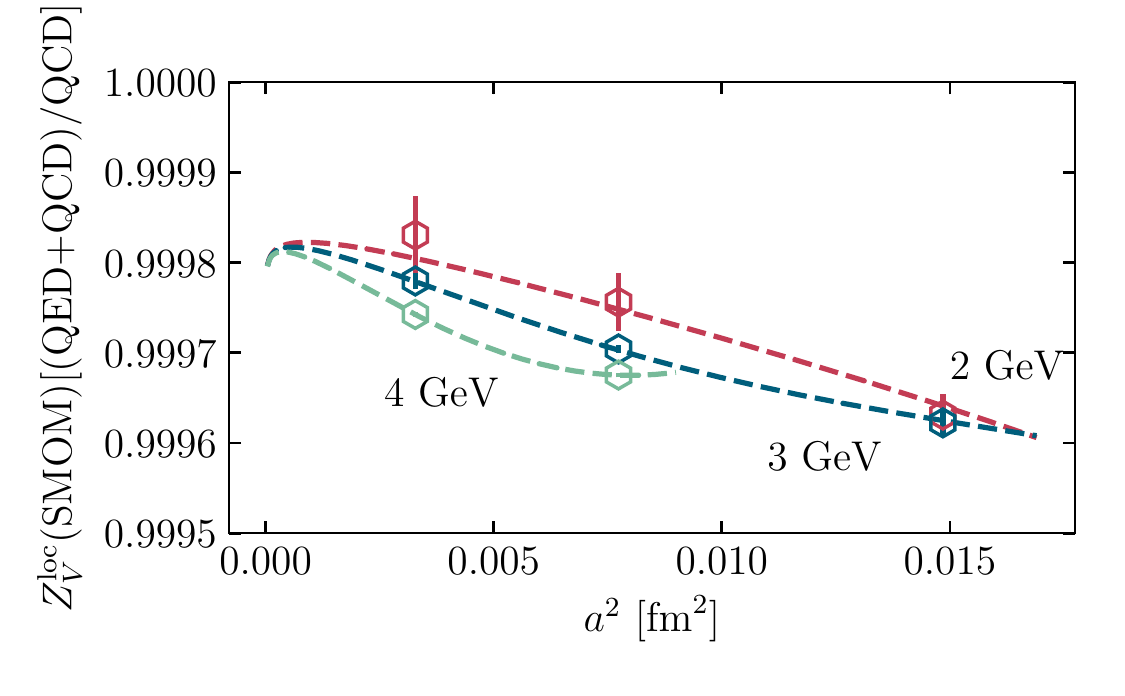}
  \caption{The ratio of $Z_V^{\text{loc}}$ values for QCD+QED to QCD 
calculated in the RI-SMOM scheme. Results are given for coarse to superfine 
lattices at $\mu$ values from 2 to 4 GeV and plotted against the square 
of the lattice spacing. The dashed lines give the result of a fit 
described in the text that shows that the results are fully described 
by a perturbative series (of which the leading coefficient is known) 
up to discretisation effects. The dips in the fit functions close to 
$a=0$ are the result of the fact that the argument of $\alpha_s$ in the fit 
function (Eq.~(\ref{eq:qedfit})) is inversely related to $a$.
}
  \label{fig:SMOM-loc-u1}
\end{figure}

Our results for the effect of quenched QED on $Z_V$ for the local HISQ current 
in the RI-SMOM scheme are given for $\mu$ values from 2 GeV to 4 GeV 
and at 3 values of the lattice spacing in Table~\ref{tab:ZV-qed-data}.  
The results are plotted in Fig.~\ref{fig:SMOM-loc-u1}.  

Given our results for the pure QCD case in Section~\ref{subsec:SMOMloc} 
we expect the results for $Z_V$ for QCD+QED to be similarly well-behaved. 
We therefore perform a fit to the ratio of $Z_V$ for QCD+QED  
to that for pure QCD 
results that allows for both discretisation effects along 
with a perturbative expansion for the ratio of renormalisation constants. 
The leading QCD effects will cancel between the numerator and 
denominator of the ratio and so the leading term in this 
expansion will be $\mathcal{O}(\alpha_{\text{QED}})$. 
We can even fix the coefficient of the leading-order term based on 
the QCD perturbation theory for the pure QCD case. The $\mathcal{O}(\alpha_s)$ 
coefficient for $Z_V^{\text{loc}}$ for pure QCD is -0.1164(3)~\cite{Chakraborty:2017hry}. 
We therefore expect that the coefficient of $\alpha_{\text{QED}}Q^2$ in the 
QED case  is $-0.1164\times 3/4 = -0.0873$. For $Q=2e/3$ this corresponds to 
an $\mathcal{O}(\alpha_{\text{QED}})$ coefficient of -0.0388. This gives 
a leading order result for $Z_V^{\text{loc}}$ of 0.9997, very close to 1.  
There will be in principle $\alpha_s\alpha_{\text{QED}}$ corrections to this 
which are likely to have an even smaller impact. 

We therefore take a fit form for the ratio of $Z_V$ values given 
in Table~\ref{tab:ZV-qed-data} of
\begin{eqnarray}
\label{eq:qedfit}
  Z_V^{\mathrm{loc}}(\mathrm{SMOM})[\mathrm{(QED+QCD)/QCD}] &=& 1 + \\ && \hspace{-10em}\alpha_{\mathrm{QED}} \left( \sum_{i } c_i \alpha_s^i (1+\sum_j d_{ij}(a\mu)^{2j} ) \right) .\nonumber
\end{eqnarray}
We use $i=0,1,2,3$ and $j=1,2,3$ fixing $c_0$ to the value 
given above. Note that $\alpha_{\text{QED}}$ does not run in this expression because 
we are using quenched QED. $\alpha_s$ in Eq.~(\ref{eq:qedfit}) is 
taken as $\alpha_{\overline{\text{MS}}}(1/a)$. 
This fit returns a $\chi^2/\text{dof}$ value of 0.25. 
The fit is plotted with the results in Fig.~\ref{fig:SMOM-loc-u1}. 

Figure~\ref{fig:SMOM-loc-u1} shows that the results for $Z_V$ behave 
as expected. The impact of quenched QED on the value of $Z_V^{\text{loc}}$ 
is tiny and indeed negligible if we imagine working to 
an accuracy of 0.1\%. Note that this follows directly 
from the analysis above in which we derive the $\mathcal{O}(\alpha_{\text{QED}})$ 
coefficient for the QCD+QED case from the pure QCD case. Because the HISQ 
action is so highly improved $Z_V^{\text{loc}}$ is very close to 1 in 
the pure QCD case. It then has to be true that the difference from 1 in $Z_V$ induced 
by QED will be over 100 times smaller than that induced by QCD. 
For the HISQ action this means that the impact of QED in $Z_V^{\text{loc}}$ 
is of order 0.03\%. This should be contrasted to the case from the 
domain-wall action where the $Z_V$ value in pure QCD is 0.7 
and so the impact 
of quenched QED is to change $Z_V$ by approximately 0.3/100 for $Q=2e/3$, 
in this case 0.3\% (see Table 6 of~\cite{Boyle:2017gzv}); this is not negligible. 

The effect of having electrically charged sea quarks would appear in 
$Z_V$ at $\mathcal{O}(\alpha_s^2\alpha_{\text{QED}})$ i.e. two orders in $\alpha_s$ below 
the leading term; the leading term comes from a photon exchange across 
a quark bubble created from a gluon. This is unlikely to change the picture significantly. 

The effect of QED on $Z_V$ is of course not a physical result and 
it needs to be combined with hadronic matrix elements for the vector 
current to understand the physical effect of QED. For this we simply take 
the values for $Z_V$ at a fixed $\mu$ value for the ensembles for which 
we have matrix element results, multiply them and extrapolate to the 
continuum limit.  Different quark 
formalisms should agree on the physical effect (on an uncharged sea). 
We will give an analysis 
of the impact of quenched QED on vector current matrix elements calculated 
with the HISQ action elsewhere. 

\section{Conclusions}
\label{sec:conclusions}

We have shown by explicit calculation how 
the vector Ward-Takahashi identity works for the HISQ action in lattice QCD. 
Renormalisation methods that make use of this identity 
will give a renormalisation constant of 1 for the conserved current as would be 
obtained in continuum QCD.  
The RI-SMOM momentum-subtraction scheme is such a scheme but the RI$'$-MOM 
scheme is not and this has implications for the accuracy achievable for 
$Z_V$ for nonconserved currents within 
each scheme.  
Our calculations have used the HISQ action but our conclusions are not specific 
to this action. 

The RI-SMOM scheme provides precise values 
for $Z_V$ for nonconserved currents 
(using momentum-sources) that are completely nonperturbative. 
Our results show that the $Z_V$ values are 
`exact'  in being free of condensate contamination. 
This means that we can simply determine $Z_V$ at a given momentum scale $\mu$ 
on a given gluon-field ensemble, multiply our vector current hadronic 
matrix element by it and then extrapolate results for the renormalised matrix 
element to the continuum limit.  
Because there is no condensate contamination there is no lower limit 
to the $\mu$ value that can be used. Statistical errors grow as 
$\mu$ is reduced but discretisation effects become smaller.  
In Section~\ref{subsec:SMOMloc} we demonstrated 
a simple method to reduce discretisation 
effects, if they are an issue, by combining 
results from two different $\mu$ values.

The RI$'$-MOM scheme can also provide precise values for $Z_V$ for nonconserved 
currents, but is not completely nonperturbative. A more critical problem with this scheme
is that the $Z_V$ values for both conserved and nonconserved currents  
have condensate contributions that begin at $1/\mu^2$. This means that 
the $Z_V$ values cannot be used to obtain accurate renormalised vector current 
matrix elements in the continuum limit without an analysis of these condensate 
contributions. This requires numbers for $Z_V$ at multiple $\mu$ values and a fit 
that includes condensate terms. If this analysis is not done, the results 
obtained in the continuum limit will be incorrect at the 1\% level.  

An alternative to the standard RI$'$-MOM scheme that avoids this problem 
is to determine $Z_V$ from a ratio of vector vertex functions for the 
conserved and nonconserved currents. We call this scheme RI$'$-MOM$_{\text{Rc}}$. 
A similarly modified RI-SMOM$_{\gamma_{\mu}}$ scheme can also be used 
to obtain an exact $Z_V$. These schemes are discussed 
in Sections~\ref{subsec:MOMloc} 
and~\ref{subsec:SMOMgloc}.  

It is straightforward to include quenched QED effects in the determination 
of the vector current renormalisation factor in a fully nonperturbative way 
using the RI-SMOM scheme and to obtain a full understanding of the 
results (including consistency with perturbation theory). 
We see only very small (below 0.1\%) effects 
for the local HISQ vector current reflecting the fact 
that the renormalisation factors in the pure QCD case 
are already very close to 1. We will include the QCD+QED $Z_V$ values in 
a future QCD+QED determination of hadronic vector current matrix elements. 

\subsection*{\bf{Acknowledgements}}

We are grateful to the MILC collaboration for the use of
their configurations and their code base.
We thank E. Follana and E. Royo-Amondarain for gauge-fixing the
superfine and fine configurations and we are grateful to E. Follana, 
S. Sharpe and A. Vladikas for useful discussions.
Computing was done on the Darwin supercomputer at the University of
Cambridge High Performance Computing Service as part of the DiRAC facility,
jointly funded by the Science and Technology Facilities Council,
the Large Facilities Capital Fund of BIS and
the Universities of Cambridge and Glasgow.
We are grateful to the Darwin support staff for assistance.
Funding for this work came from the
Science and Technology Facilities Council
and the National Science Foundation. 

\appendix

\section{HISQ conserved current} \label{app:cons-curr}

The forward HISQ conserved current corresponding to the simple backward finite 
difference operator $\Delta^{\mu,-}$ in Eq.~(\ref{eq:ward-pos}) is given by
\begin{align}
\label{eq:Jcons}
  &J^{\mu,+}(\tilde{x}) = \frac{1}{2}[\overline{\psi}(x) \gamma_{\mu} W_{\mu}(x)\psi(x+\hat{\mu}) + h.c.] \\
  &+ \frac{1}{16}[\overline{\psi}(x) \gamma_{\mu} X_{\mu}(x)\psi(x+\hat{\mu}) + h.c.] \nonumber \\
  &- \frac{1}{48}[\overline{\psi}(x-2\hat{\mu})\gamma_{\mu}X_{\mu}(x-2\hat{\mu})X_{\mu}(x-\hat{\mu})X_{\mu}(x)\psi(x+\hat{\mu}) \nonumber \\
  &+ \overline{\psi}(x-\hat{\mu})\gamma_{\mu}X_{\mu}(x-\hat{\mu})X_{\mu}(x)X_{\mu}(x+\hat{\mu})\psi(x+2\hat{\mu}) \nonumber \\
  &+ \overline{\psi}(x)\gamma_{\mu}X_{\mu}(x)X_{\mu}(x+\hat{\mu})X_{\mu}(x+2\hat{\mu})\psi(x+3\hat{\mu}) + h.c.] \nonumber
\end{align}
where $W$ are HISQ links and $X$ are the links after the first level of HISQ smearing in the notation
of \cite{Follana:2006rc}. Note that $J^{\mu,+}$ sits on the link between 
$x$ and $x+\hat{\mu}$; $\tilde{x}$ is the halfway point on that link.  
The backward conserved current $J^{\mu,-}$ is the same but with $x \to x-\hat{\mu}$ and
$x + \hat{\mu} \to x$.
More complicated conserved currents can be defined in conjunction with higher-order 
difference operators for $\Delta^{\mu,\pm}$ but we do not do that here. 

\section{Renormalisation of the 1-link vector current} \label{app:1link}

\begin{table}
\caption{
Column 2 gives the tadpole-improvement factor $u_0$ used in the definition of 
the 1-link current (Eq.~(\ref{eq:1linkfwddef})). This is the mean value of the gluon 
field $U_{\mu}$ in Landau gauge.
Column 3 gives the results for the $Z_V$ values determined from the form factor 
using the matrix element of the temporal 1-link current between two pions at 
rest\footnote{We thank J. Simone for providing the $u_0$ values 
and J. Koponen and A.C. Zimermmane-Santos for providing the $Z_V(\text{F(0)})$ 
values.}. 
The asterisk next to set 6 is to denote that the results given here are actually 
for another fine ensemble with $am_l=0.0074$, $am_s=0.037$ and $am_c=0.44$.
}
\label{tab:u0}
\begin{tabular}{lll}
Set & $u_0$ & $Z_V^{\mathrm{1link}}(\mathrm{F(0)})$ \\
\hline
1 & 0.820192(14) & 1.0332(23) \\
2 & 0.834613(14) & 1.0307(7) \\
6* & 0.852477(9) & 1.0193(9) \\
7 & 0.870935(5) & 1.0064(28) \\
\end{tabular}
\end{table}

\begin{table}
\caption{Renormalisation factors for the (forward) 1-link HISQ vector current for a variety 
of $\mu$ values (given in column 2) on gluon field configurations at different lattice 
spacing values (denoted by the set number in column 1). Column 3 gives results using the 
RI-SMOM scheme. 
}
\label{tab:1-link}
\begin{tabular}{lll}
Set & $\mu$ [GeV] & $Z_V^{\mathrm{1link}}(\mathrm{SMOM})$ \\
\hline
1 & 1 & 0.9617(11) \\
2 & 1 & 0.9713(19) \\
\hline
1 & 2 & 0.93516(16) \\
2 & 2 & 0.94966(20) \\
6 & 2 & 0.96695(11) \\
7 & 2 & 0.97996(34) \\
\hline
2 & 2.5 & 0.94236(11) \\
\hline
2 & 3 & 0.939193(87) \\
6 & 3 & 0.954643(37) \\
7 & 3 & 0.97225(12)  \\
\hline
6 & 4 & 0.948641(20)  \\
7 & 4 & 0.965353(56)  \\
\end{tabular}
\end{table}

\begin{figure}
  \includegraphics[width=0.47\textwidth]{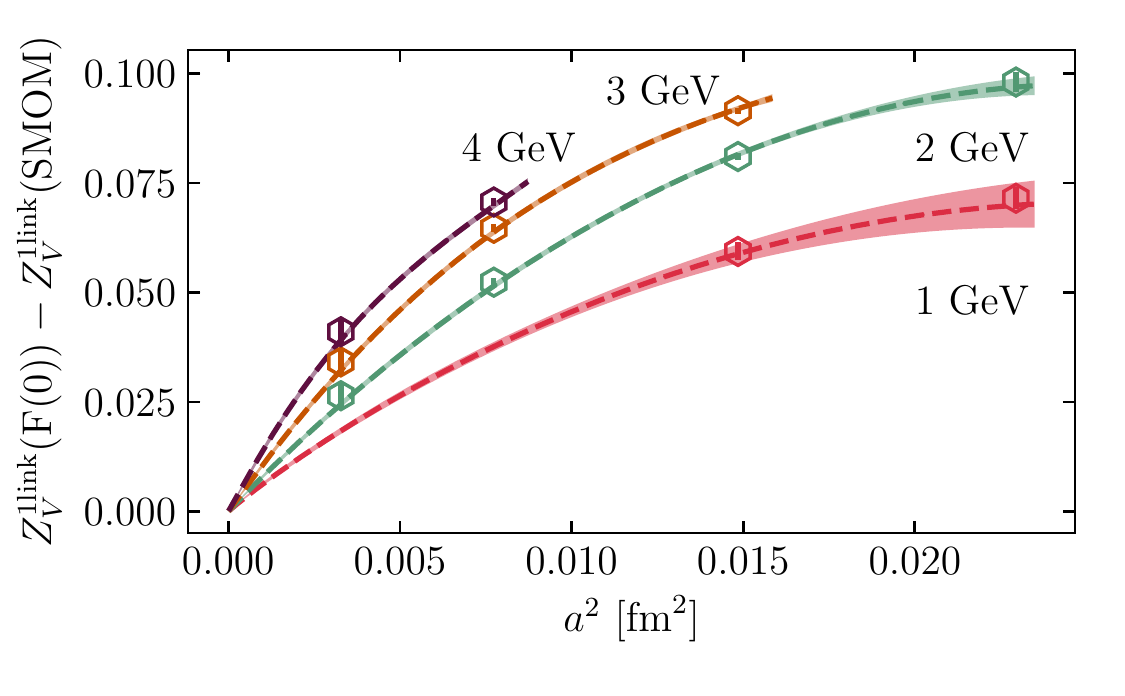}
  \caption{$Z_V^{\mathrm{1link}}(\mathrm{SMOM})$ for $\mu$ values between 1 GeV and 4 GeV, 
plotted as a difference to the corresponding $Z_V$ at that lattice spacing 
obtained from the vector form factor at 
zero momentum-transfer. The fit shown (see Eq.~(\ref{eq:1linkdifffit})) accounts for
  discretisation effects only.  
  }
  \label{fig:1linkdiff}
\end{figure}

Quark-line disconnected contributions for vector current-current correlators 
require the use of a taste-singlet vector current for staggered quarks. 
This has the same taste as the conserved current but 
it is often more convenient 
to use a simpler current than the conserved one. Here we discuss the renormalisation 
of the nonconserved 1-link point-split vector current using 
momentum-subtraction schemes. The qualitative picture is the same as that 
for the local current and so we simply include RI-SMOM results in this Appendix 
for completeness. They are relevant to our ongoing calculations 
of, for example, the quark-line disconnected pieces of the hadronic 
vacuum polarisation contribution to the anomalous magnetic moment of 
the muon. 

We consider the 1-link point-split vector current with 
spin-taste ${(\gamma_{\mu} \otimes I)}$. 
The operators that 
we use include gluon fields between the point-split quark fields 
to maintain gauge invariance.  We take these gluon fields to 
be `thin links' i.e. no smearing is applied to them. 
We considered the two simplest
constructions of this current. 
One, which we denote the forward 1-link operator, 
is the conserved current with all 3-link
terms removed:
\begin{equation}
\label{eq:1linkfwddef}
  j^{\mathrm{1link}}_{\mu} \equiv \frac{1}{2u_0}\overline{\psi}(x) \overline{(\gamma_{\mu} \otimes 1)} U_{\mu}(x) \psi(x+\hat{\mu}) + h.c. 
\end{equation}
The other 1-link operator we consider is the symmetric operator 
\begin{eqnarray}
\label{eq:1linksymdef}
  j^{\mathrm{1link\mhyphen symm}}_{\mu} &\equiv& \frac{1}{2u_0}\overline{\psi}(x) \overline{(\gamma_{\mu} \otimes 1)} U_{\mu}(x) \psi(x+\hat{\mu}) \\
  &+& \frac{1}{2u_0}\overline{\psi}(x) \overline{(\gamma_{\mu} \otimes 1)} U^{\dagger}_{\mu}(x-\hat{\mu}) \psi(x-\hat{\mu}) . \nonumber
\end{eqnarray}
The two definitions coincide with the MOM kinematics. 
In the SMOM case, while the values produced from the two different definitions
are not identical they agree within our statistical 
uncertainties. In what follows we then only present results 
for the forward 1-link current. 

Note that in the definitions of the 1-link current above we include a factor $1/u_0$. 
$u_0$ is a `tadpole-improvement' factor~\cite{Lepage:1992xa} which can be used, 
as here, to reduce the mismatch between lattice currents containing gluon fields 
and their continuum counterparts. $u_0$ works by cancelling universal effects from 
tadpole diagrams that arise from the construction of the lattice gluon field. 	
$u_0$ can in principle be any suitable ensemble average of a function of the gluon 
field that achieves this. Here we use the mean value of the gluon
field $U_{\mu}$ in Landau gauge as the most appropriate form of $u_0$ in this case.  
The values for $u_0$ depend on the ensemble and are listed in Table~\ref{tab:u0} 
\footnote{$Z_V$ for the tadpole-improved current is $u_0$ times $Z_V$ for the 
current with no tadpole-improvement.}. 

We proceed for the 1-link case in the same way as for the cases discussed in the main body 
of the paper. The wavefunction renormalisation is exactly the same as before. We 
calculate the vertex function for the 1-link current using an appropriate 
projector. 
For the RI-SMOM case we use
\begin{equation}
\label{eq:zv1linkdefhisq}
\frac{Z_q(q)}{Z_V(q)} = \frac{i}{48V^{\text{1link}}_{\gamma\otimes I}} \sum_{\mu,\nu} (-ia\hat{q}_{\mu})\frac{a\hat{q}_{\nu}}{(a\hat{q})^2}\Tr\left[ \overline{\overline{(\gamma_{\nu}\otimes I)}}\Lambda^{\mu}_V\right] . 
\end{equation}

In determining $Z_V^{\text{1link}}$ an additional technical detail for 
point-split operators is that we must divide the vertex function in the 
full theory by the result of the tree-level (noninteracting) case. 
This was discussed previously for the conserved 
current in the RI$'$-MOM case in Section~\ref{subsec:MOMcons} (and denoted 
$V_{\gamma\otimes I}$ in Eq.~(\ref{eq:zvmomdefhisq}) and above). 
The tree-level result for the forward 1-link current for the RI-SMOM
kinematics is: 
\begin{equation}
\label{eq:tree1link}
V^{\text{1link}}_{\gamma\otimes I}(\text{SMOM}) =  \frac{1}{2} \prod_{\mu}(e^{iap_{2,\mu}(S-T)_{\mu}} + e^{-iap_{1,\mu}(S-T)_{\mu}}) \\
\end{equation}

A further technical detail arises when using twisted boundary 
conditions to insert momentum with point-split operators
in the vertex functions that we calculate. The propagator with 
twisted momentum can be written in terms of the untwisted one as
\begin{equation}
  \tilde{S}(x,p) = e^{-i\theta x}S(x,p+\theta) .
\end{equation}
We want the vertex function for a point-split operator to take the following 
form (using a 1-link operator $\Gamma$ as an example, but dropping the gluon fields for clarity):
\begin{align}
  &\sum_x \gamma_{5} e^{i(p_1 + \theta_1)x} S^{\dagger}(x,p_1 + \theta_1) \\ &\times \gamma_5 \Gamma_{\mu} e^{-i(p_2+\theta_2)x} S(x+\hat{\mu},p_2+\theta_2) \nonumber \\
  &= \sum_x \gamma_5 e^{ip_1x}\tilde{S}^{\dagger}(x,p_1) \gamma_5 \Gamma_{\mu} e^{-ip_2x} e^{ia\theta_{2,\mu}} \tilde{S}(x+\hat{\mu},p_2) . \nonumber
\end{align}
The factor $e^{ia\theta_{2,\mu}}$ has to be inserted by hand. 

Our results for $Z_V^{\text{1link}}(\text{SMOM})$ using the RI-SMOM scheme 
are given in Table~\ref{tab:1-link} for a variety of 
$\mu$ values for three values of the lattice spacing.
We expect the $Z_V$ values obtained with RI-SMOM to be well-behaved 
and free of condensate contributions because of the protection of the 
Ward-Takahashi identity, as for the local current discussed in 
Section~\ref{subsec:SMOMloc}. We can test this, as was done for the local 
case, by taking a difference of the $Z_V$ values with those obtained 
from the form factor method. 

The results for $Z_V$ from the form factor method are given in Table~\ref{tab:u0} 
for a variety of $\mu$ values and on ensembles with a range of 
lattice spacing values. 
The results for the difference of $Z_V$ values between the form 
factor and RI-SMOM methods is plotted in Fig.~\ref{fig:1linkdiff}. 
We show the results of a simple fit to a sum of possible discretisation 
effects: 
\begin{eqnarray}
\label{eq:1linkdifffit}
\Delta Z_V^{\text{1link}}(a,\mu) &=& \sum_{i=0,j=1}^{2,3} c_{ij} \alpha^i_s (a\mu/\pi)^{2j} \nonumber \\
&+& \sum_{i=0,j=1}^{2,3} d_{ij}\alpha^i_s (a\Lambda/\pi)^{2j} \,.
\end{eqnarray}
Here $\alpha_s$ is in the $\overline{\text{MS}}$ scheme at scale $1/a$. 
We have to include $(a\Lambda)^n$ terms as well as $(a\mu)^n$ terms 
here because of the relatively large discretisation effects in the 
$Z_V$ values obtained from the form factor method. 
The priors on the coefficients of the fit are taken as: $0\pm 3$.  
The fit gives a $\chi^2/\text{dof}$ = 0.9. This confirms that, 
again in this case, the RI-SMOM method gives a well-behaved result for $Z_V$. 
 
The conclusion from this is that the renormalisation factors for 
the 1-link current obtained in the RI-SMOM scheme on the lattice can 
be used straightforwardly, and in a fully nonperturbative way, to 
renormalise matrix elements of the 1-link current obtained in a lattice calculation. 
This means that values can be taken, for example from Table~\ref{tab:1-link}, 
for a fixed $\mu$ value on each ensemble. The $\mu$ chosen can take any value 
and the only limitation on taking it to have a small value (for minimal 
discretisation effects) is that of the 
statistical errors that grow as $\mu$ is reduced. 

\bibliography{zvpaper}

\end{document}